\documentclass[11pt,a4paper]{article}
\usepackage{longtable}
\usepackage{amsmath}
\usepackage{amssymb}
\usepackage{graphicx}
\usepackage{bm}
\usepackage{footmisc}
\usepackage{afterpage}
\usepackage{float}
\usepackage{lscape}
\usepackage{longtable}
\usepackage{float}
\usepackage{slashed}
\usepackage{morefloats}
\usepackage{afterpage}

\newcommand*{\no}{\noindent}
\newcommand*{\bea}{\begin{eqnarray}}
\newcommand*{\eea}{\end{eqnarray}}
\newcommand*{\be}{\begin{equation}}
\newcommand*{\ee}{\end{equation}}

\newcommand*{\pref}[1]{(\ref{#1})}

\newcommand*{\mn}{{\mu\nu}}

\newcommand*{\prefr}[2]{(\ref{#1}-\ref{#2})} 
\newcommand*{\nn}{\nonumber}

\newcommand*{\tr}{\mathrm{tr}}

\newcommand{\bma}{\begin{pmatrix}}
\newcommand{\ema}{\end{pmatrix}}

\newcommand*{\la}{\left\langle}
\newcommand*{\ra}{\right\rangle}

\usepackage[square,comma,sort&compress,numbers]{natbib}

\title{The quenched SU(2) scalar-gluon vertex in minimal Landau gauge}

\author{Axel Maas\\
Institute of Physics, NAWI Graz, University of Graz,\\
Universit\"atsplatz 5, A-8010 Graz, Austria}

\begin{document}

\maketitle

\begin{abstract}

The question of whether confining effects are visible in correlation functions is a long-standing one. Complementing investigations on the propagators of fundamental and adjoint scalar matter particles here the quenched scalar-gluon vertex is investigated. For this purpose a multitude of lattice setups in two, three, and four dimensions is analyzed in quenched SU(2) lattice gauge theory. Though both cases are quantitatively different, neither a qualitative difference nor any singularities are observed.

\end{abstract}

\section{Introduction}

It is a long-standing question, whether any effects of confinement manifest themselves in correlation functions involving a finite number of fields. In particular, whether the differences between fundamental matter, associated with the Wilson confinement criterion, and adjoint matter, which is always affected by string-breaking, should be visible \cite{Alkofer:2006fu}. At various times, this has been associated with anything from non-trivial analytic structures to various types of singularities, both of propagators and vertices, see \cite{Alkofer:2000wg,Maas:2011se,Fischer:2006ub,Binosi:2009qm,Boucaud:2011ug,Vandersickel:2012tg,Roberts:2015lja,Huber:2018ned} for reviews.

Particularly interesting in this respect are quenched SU($N$) gauge theories, where fundamental and adjoint matter manifestly differ. To disentangle confining effects from those of chiral symmetry it is furthermore useful to switch to scalar matter, as it is not expected that this will change this question.

At the level of the propagators, this has been studied in \cite{Fister:2010yw,Macher:2011ad,Capri:2012ah,Capri:2013oja,Hopfer:2013via,Maas:2013aia,Capri:2017abz,Maas:2016edk,Maas:2018sqz}. No qualitative distinction between the fundamental and the adjoint case have been found, though quantitatively both differ substantially. However, in both cases dynamical mass, or at least scale, generation was observed, which is often associated with a gaping of the degrees of freedom. Finally, both cases showed substantial indications for violations of positivity, which at least guarantees their absence as physical degrees of freedom. Unfortunately, also no indication of singularities or other obvious qualitatively remarkable features were observed, which could be associated to a confining force, however it is defined.

On the other hand, it has been argued that confinement may be an interaction effect, e.\ g.\ due to singularities in the vertex interaction, and not manifest in the propagators themselves \cite{Fister:2010yw,Zwanziger:2010iz,Alkofer:2006gz,Alkofer:2008tt,Schwenzer:2008vt}. This motivates the present study, where the quenched scalar-gluon vertex for SU(2) will be studied in two, three, and four dimensions. The advantage, in comparison to the quark-gluon vertex \cite{Alkofer:2006gz,Alkofer:2008tt,Kizilersu:2006et,Sternbeck:2017ntv,Binosi:2016wcx,Huber:2018ned}, is that there is only a single form factor involved, and again chiral symmetry is not an issue. Studying lower dimensions, besides allowing to reach much deeper into the infrared, allows to compare to cases were geometric Wilson confinement already in QED arises, and where no dynamics occur in the gauge sector. This allows to systematically switch on and off various contributions.

The technical setup of the employed lattice calculations will be discussed in section \ref{s:tech} and appendix \ref{a:ls}. It follows closely \cite{Maas:2016edk,Maas:2018sqz,Cucchieri:2006tf}, utilizing that the scalar-gluon vertex is quite similar to the comparatively well-studied ghost-gluon vertex \cite{Cucchieri:2006tf,Cucchieri:2008qm,Maas:2007uv,Cucchieri:2004sq,Sternbeck:2005re,Sternbeck:2012mf,Zafeiropoulos:2019flq}. Renormalization is discussed in section \ref{s:ren}. In this context also the discretization artifacts need to be discussed in more detail, as they are, quite similarly to the propagator case \cite{Maas:2016edk,Maas:2018sqz}, much stronger for the adjoint case than for the fundamental case.

The results are finally presented in section \ref{s:res}. Unfortunately, as will be emphasized in the concluding section \ref{s:con}, no qualitative distinction is found for the fundamental and the adjoint case. Though quantitatively both cases are quite distinct, with the larger modifications compared to the tree-level case for the adjoint scalar. This is just as for the propagators \cite{Maas:2016edk,Maas:2018sqz}. Also, no singular behavior is found for any of the momentum configurations studied here.

Some preliminary results were available in \cite{Maas:2011yx}, and are superseded by the present work.

\section{Technical setup}\label{s:tech}

The calculations follow closely the one for the ghost-gluon vertex in \cite{Cucchieri:2006tf}, as both vertices are very similar in technical respect. The quenched gauge-fixed configurations are obtained as described in \cite{Cucchieri:2006tf}, i.\ e.\ using the Wilson gauge action and creating decorrelated configurations using a mix of heatbath and overrelaxation sweeps. A full list of the lattice settings and number of configurations can be found in table \ref{tcgf} in appendix \ref{a:ls}.

The decorrelated configurations are then gauge-fixed to minimal Landau gauge using an adaptive stochastic overrelaxation method \cite{Cucchieri:2006tf}. Minimal Landau gauge corresponds to an average with flat weight over all Gribov copies in the first Gribov region, i.\ e.\ those Gribov copies with positive semi-definite Faddeev-Popov operator \cite{Maas:2011se}. The employed algorithms appears to implement this prescription faithfully, as far as has been investigated \cite{Maas:2015nva}. However, at least the fundamental scalar propagator in the unquenched theory shows essentially no dependence on how Gribov copies are treated, in contrast to the gauge propagators \cite{Maas:2010nc}. In addition, the only vertex whose has been investigated with respect to the influence of Gribov copies, the ghost-gluon vertex \cite{Sternbeck:2012mf}, did not show a stronger dependence than the corresponding propagators. It thus seems to be plausible that the treatment of Gribov copies will have only a minor, or even negligible, impact on the results presented here.

The scalar-gluon vertex has, just like the ghost-gluon vertex, only a single transverse form factor for SU(2). Any longitudinal form factor is inaccessible on the lattice, because only non-amputated correlation functions can be determined. Following \cite{Cucchieri:2004sq,Cucchieri:2006tf,Maas:2013aia}, the corresponding form-factor is extracted by
\be
G_u=\frac{\Gamma^\text{tl}_{\mu aij}\la A_\mu^a \Delta_{ij}^{-1}\ra}{\Gamma^\text{tl}_{\mu bkl}D_\mn^{bc}D^{km}D^{ln}\Gamma^\text{tl}_{\nu cmn}}\label{vertex}
\ee
\no where $\Gamma^\text {tl}$ is the lattice tree-level vertex \cite{Maas:2013aia}\footnote{Note that in the dynamical calculations of \cite{Maas:2013aia}, care had to be taken because of the unbroken custodial symmetry, which would yield a vanishing naive vertex. In the present quenched case this is not necessary.},
\be
\Gamma^\text{tl}_{\mu aij}(k,p,q)=\frac{iga}{6}\tau^a_{ij}\sin\left(\frac{\pi}{N}(P-Q)_\mu\right)\cos\left(\frac{\pi}{N}(P+Q)_\mu\right)\nn,
\ee
\no where $P$ and $Q$ are the integer-valued lattice momenta and $N$ is the extension of the lattice. The gluon propagator $D_\mn$ and corresponding respective scalar fundamental and adjoint propagators $D$ appear to amputate the correlation function to end up with the final unrenormalized vertex form factor $G_u$.

The appearing inverse covariant Laplacian $\Delta$ in \pref{vertex} arises from integrating out the scalar field in the quenched case. Just as in \cite{Maas:2016edk,Maas:2018sqz}, the simplest lattice discretizations \cite{Greensite:2006ns} $\Delta_L$ of the fundamental
\be
-\Delta^2_L=-\sum_\mu\left(U_\mu(x)\delta_{y(x+e_\mu)}+U_\mu^\dagger(x-\mu)\delta_{y(x-e_\mu)}-2\delta_{xy}\right)+m_0^2\delta_{xy}\label{covf},
\ee
and adjoint
\bea
-\Delta^2_L&=&-\sum_\mu\left(U^a_\mu(x)\delta_{y(x+e_\mu)}+U_\mu^{a\dagger}(x-\mu)\delta_{y(x-e_\mu)}-2\delta_{xy}\right)+m_0^2\delta_{xy}\nn\\
U_{\mu bc}^a&=&\frac{1}{2}\tr\left(\tau^b U_\mu^\dagger\tau^c U_\mu\right)\label{cova},
\eea
Laplacians are used. Their inversion is performed, as for the propagators \cite{Maas:2016edk,Maas:2018sqz,Maas:2010nc}, using a conjugate gradient algorithm with explicit exclusion of the zero momentum case.

At any rate, both zero and maximal lattice momenta yield a vanishing denominator in \pref{vertex}, and are thus inaccessible in the present calculation. The parameter $m_0$ provides the tree-level mass of the scalars in lattice units. Here, the cases $m=m_0/a=0$, $0.1$, $1$, and $10$ GeV are investigated.

The form factor is a function of three momenta, $G_u(p^2,q^2,k^2)$, where $p^2$ is the gluon momentum, $q^2$ is the first scalar momentum, and $k^2$ the second one. For the fundamental case it is the one of the anti-scalar. Following the setup for other vertices from \cite{Cucchieri:2006tf}, three different momentum configurations are investigated: The soft gluon or  back-to-back case $p^2=0$ with $q^2=k^2$, the equal or symmetric case $p^2=q^2=k^2$, and the orthogonal case $p^2=q^2$ and $pq=0$ implying $k^2=(p+q)^2=p^2+q^2$. These are implemented for the integer lattice momenta, which implies that for the symmetric case the three momenta cannot be arranged within two dimensions, and therefore this configuration is impossible in two dimensions. All results will be given in terms of the physical momenta rather than lattice momenta.

Note that an inversion is necessary for every scalar momentum, and thus for these momentum configurations in total $2N$ for each field configuration. The computation time for the inversions scales at least like $N^5$, and thus the total computation time at least like $2N^6$. In addition, quite precise data is needed to see the systematic trends, and thus a substantial amount of statistics. This gives as a lower bound to the computation time of $2\times 10^3 N^6$ in units of the time necessary for the smallest possible lattice, substantially limiting the accessible lattice volumes.

\section{Renormalization}\label{s:ren}

Just like the other three-point vertices in Landau gauge the scalar-gluon vertex does renormalize trivially, though a finite renormalization is possible. It is important to note that this statement applies to the form-factor. Thus, even though the scalar propagator renormalizes non-trivially in \pref{vertex} the unrenormalized scalar propagators needs to be used to remove the external propagators from the full correlation function. Thus, at most a multiplicative renormalization of $G_u$ is necessary, at least perturbatively. Just like for the propagators \cite{Maas:2016edk,Maas:2018sqz}, it is found that the perturbative renormalization is sufficient also non-perturbatively. In addition, lattice spacing effects, vanishing for $a\to 0$, can affect the form factor. If these are momentum-independent, they can also be counteracted by renormalization.

Thus, a multiplicative renormalization condition is applied to the form factor
\be
G(\mu^2,\mu^2,4\mu^2)=Z_GG_u(\mu^2,\mu^2,2\mu^2)=1\label{rcond},
\ee
\no with $\mu=1.5$ GeV, $Z_G$ the corresponding renormalization constant, and $G$ the renormalized form factor. This choice allows the same condition in two, three, and four dimensions, and for all lattice spacings equally. Using the more conventional symmetric configuration would have required a different treatment in two dimensions. Using alternatively the back-to-back momentum configuration was not done because of possible infrared singularities \cite{Fister:2010yw}, though none were ultimately encountered. The relatively low value of $\mu$ was chosen to have even on the coarsest lattice still the same renormalization point and being not too close to the largest momenta. The actual value of $Z_G$ was obtained by linear interpolation between the two momenta closest to $\mu$, and the error in its determination propagated to the renormalized form factor.

In the fundamental case, the value of $Z_G$ is always, within about 2$\sigma$ statistical error, independent of the lattice spacing, and within a few percent of $Z_G=1$. Though some systematic trend seems to be present, this would require probably one to two orders of magnitude more configurations to clarify by obtaining $Z_G$ at the per mille level. For the present purpose, $Z_G$ can thus be assumed to be essentially independent of the lattice spacing, though the results shown in section \ref{s:res} will still be renormalized according to \pref{rcond}.

\begin{figure}
\includegraphics[width=0.5\linewidth]{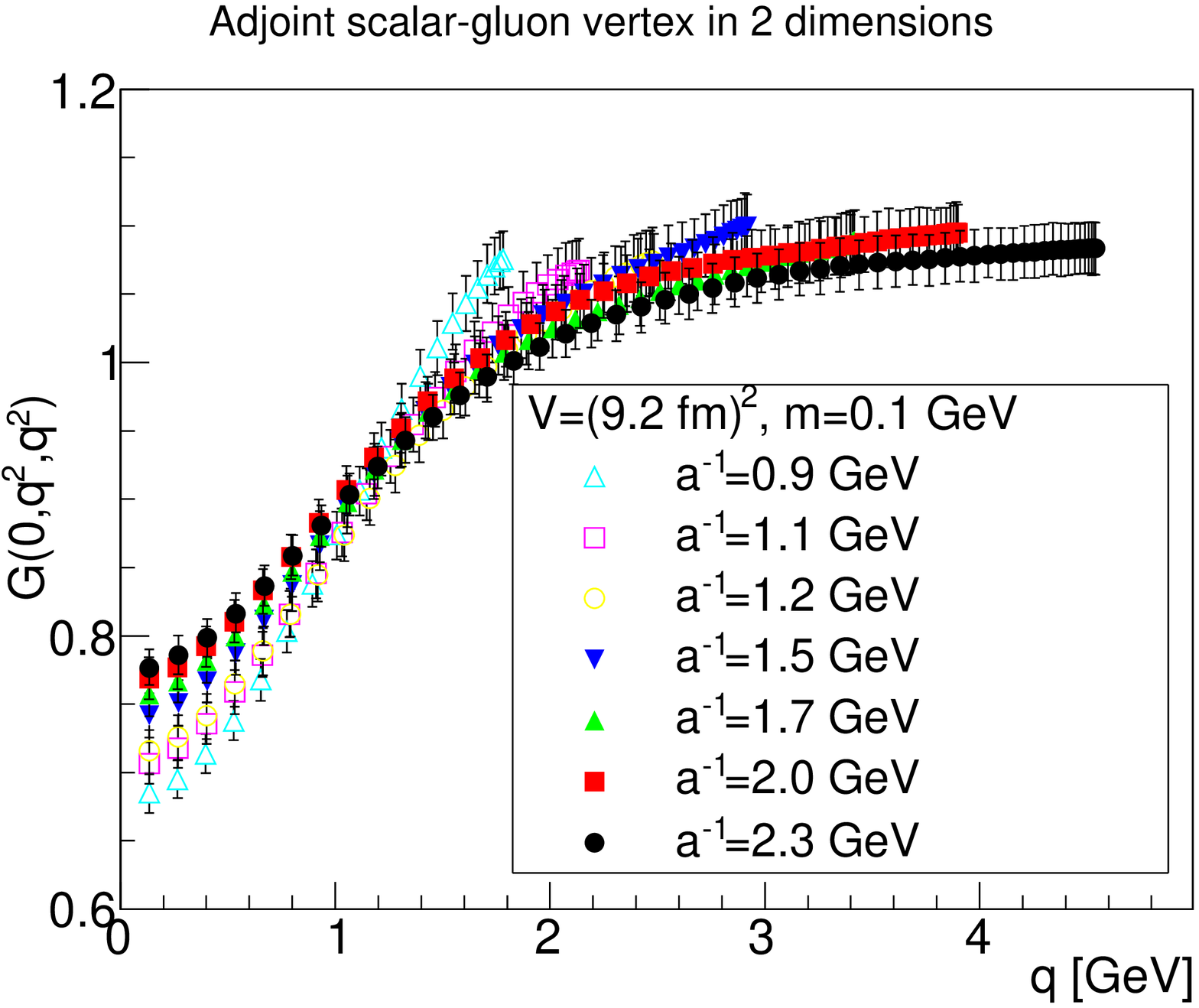}\includegraphics[width=0.5\linewidth]{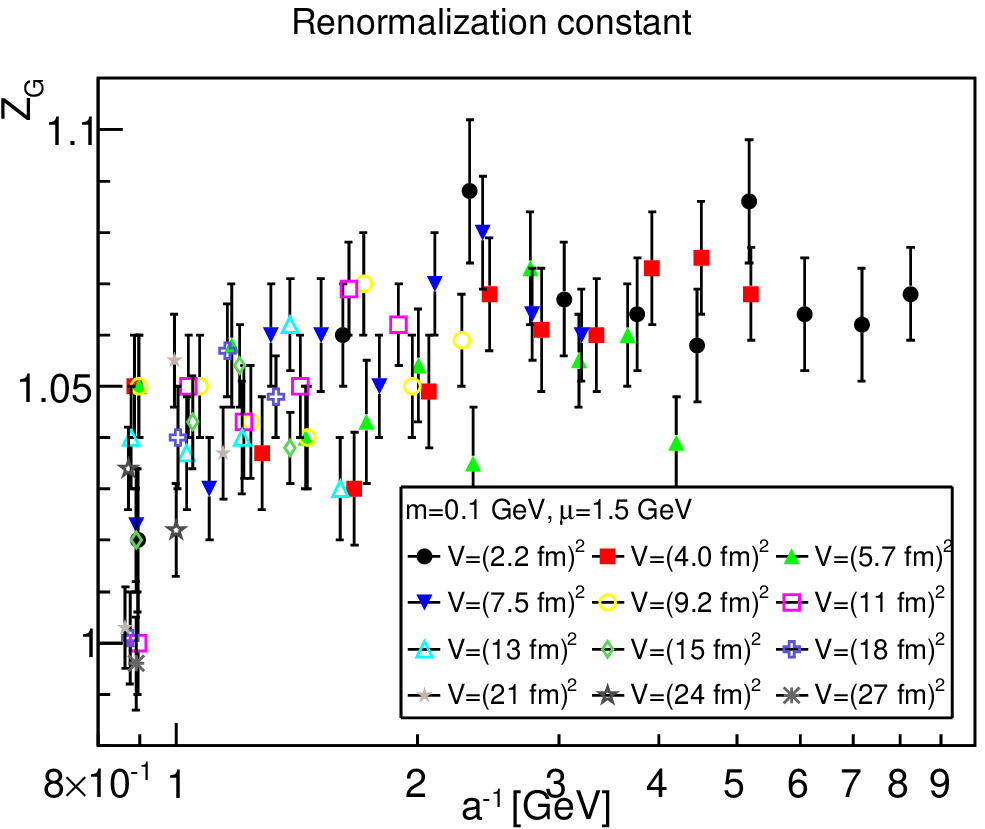}\\
\includegraphics[width=0.5\linewidth]{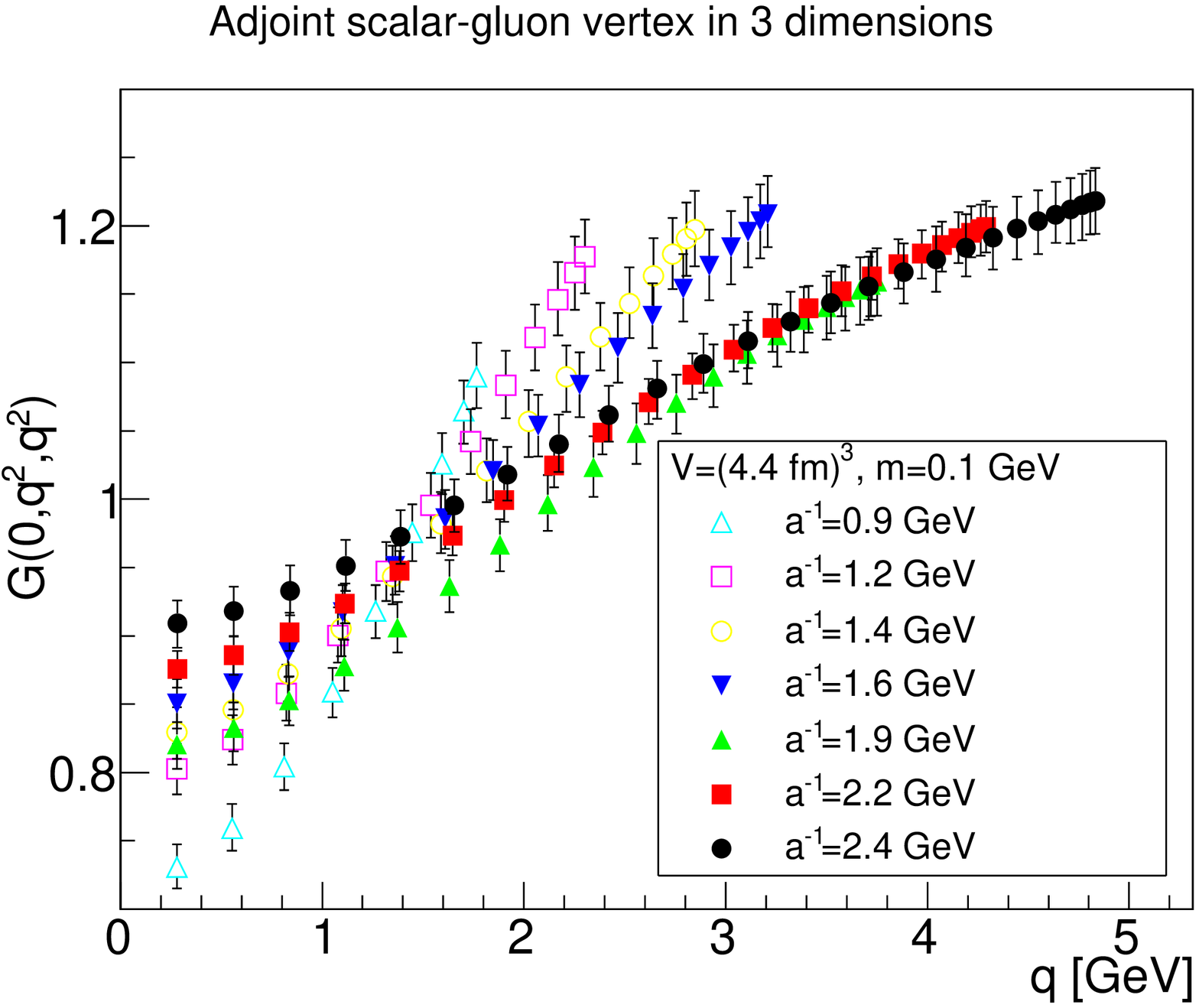}\includegraphics[width=0.5\linewidth]{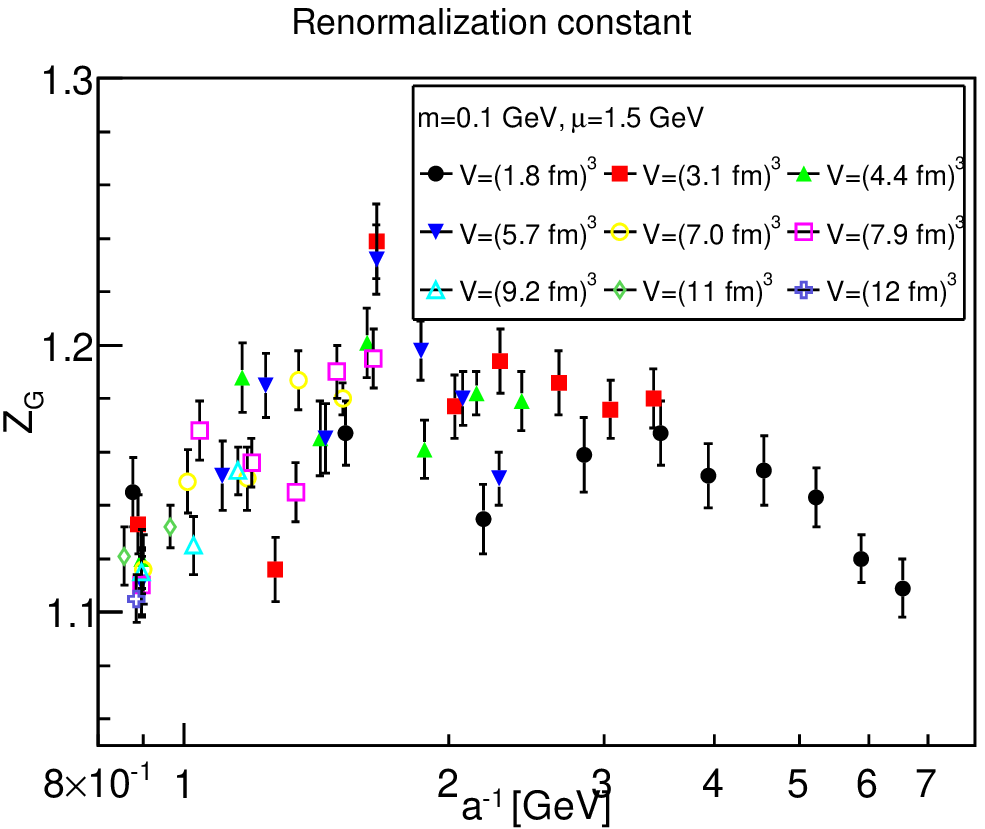}\\
\includegraphics[width=0.5\linewidth]{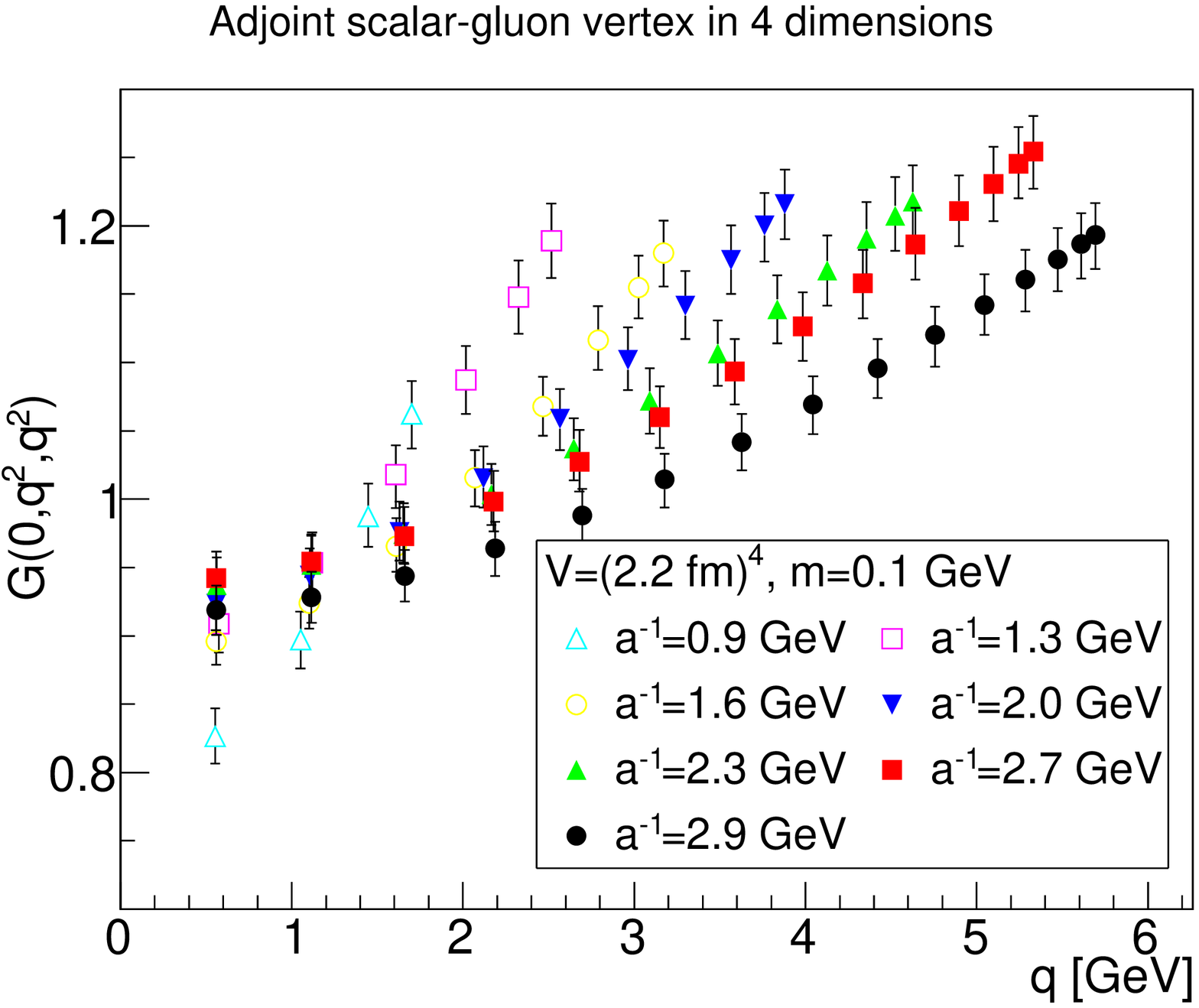}\includegraphics[width=0.5\linewidth]{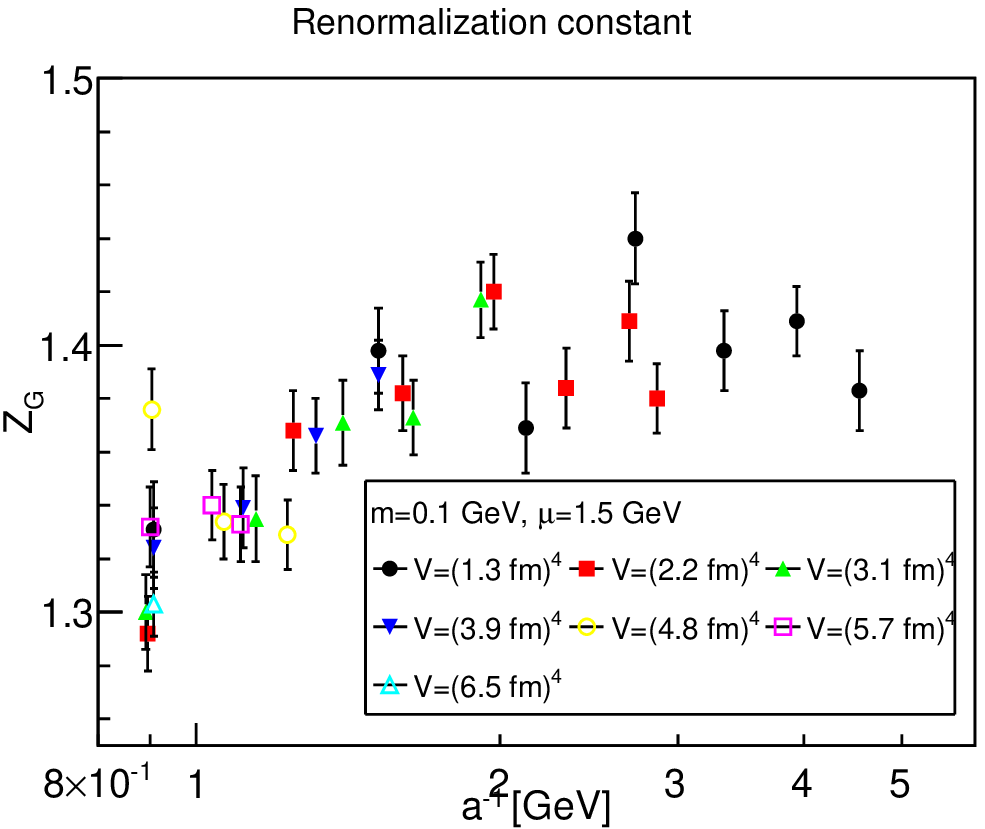}
\caption{\label{fig:adep}The discretization dependence of the vertex in the back-to-back configuration at fixed volume (left panels) and $Z_G$ (right panels) for $m=0.1$ GeV in two (top panels), three (middle panels), and four (bottom panels) dimension.}
\end{figure}

The situation is different in the adjoint case. This is because the vertex is stronger affected by momentum-dependent discretization artifacts than in the fundamental case, just as for the propagator \cite{Maas:2018sqz}. These discretization artifacts mix in with the renormalization. This behavior is exemplified in figure \ref{fig:adep}, where the vertex is shown for different discretizations at fixed volume alongside $Z_G$ for $m=0.1$ GeV. In two and three dimensions it is seen how in the ultraviolet the results start to agree above roughly $a^{-1}=1.75$ GeV. The differences in the infrared stay even above this discretization in three dimensions, and even at $a^{-1}=2.4$ GeV no convergence is visible. The situation is even worse in four dimensions, where even for $a^{-1}>2.7$ GeV no convergence is seen in the ultraviolet. Instead, there appears now to be convergence in the infrared. Still, the effect should not be overstated, as it does affect the results substantially, but rather at the ten percent level. Nonetheless, this gives reason enough to study the adjoint vertex below at fixed discretizations, and not just compare different volumes at the finest available discretizations, as is possible in the fundamental case.

\section{Results}\label{s:res}

One result, which is almost independent of dimensionality and representation, is that the vertex for a tree-level mass of $m=10$ GeV is compatible within errors with one, i.\ e.\ it remains tree-level-like within errors. The only exception is the four-dimensional adjoint case. There, a slight modification of a few percent is visible, barely above the statistical precision. This behavior is a strongly attenuated version of what is seen in the 1 GeV case, a slight increase at large momenta. Furthermore, the results for the vertices for tree-level masses $m=0$ GeV and $m=0.1$ GeV are, within statistical errors, identical. Thus, in the following only results for tree-level masses $m=0$ GeV and $m=1$ GeV will be shown explicitly.

\subsection{Fundamental case}

\begin{figure}
\includegraphics[width=\linewidth]{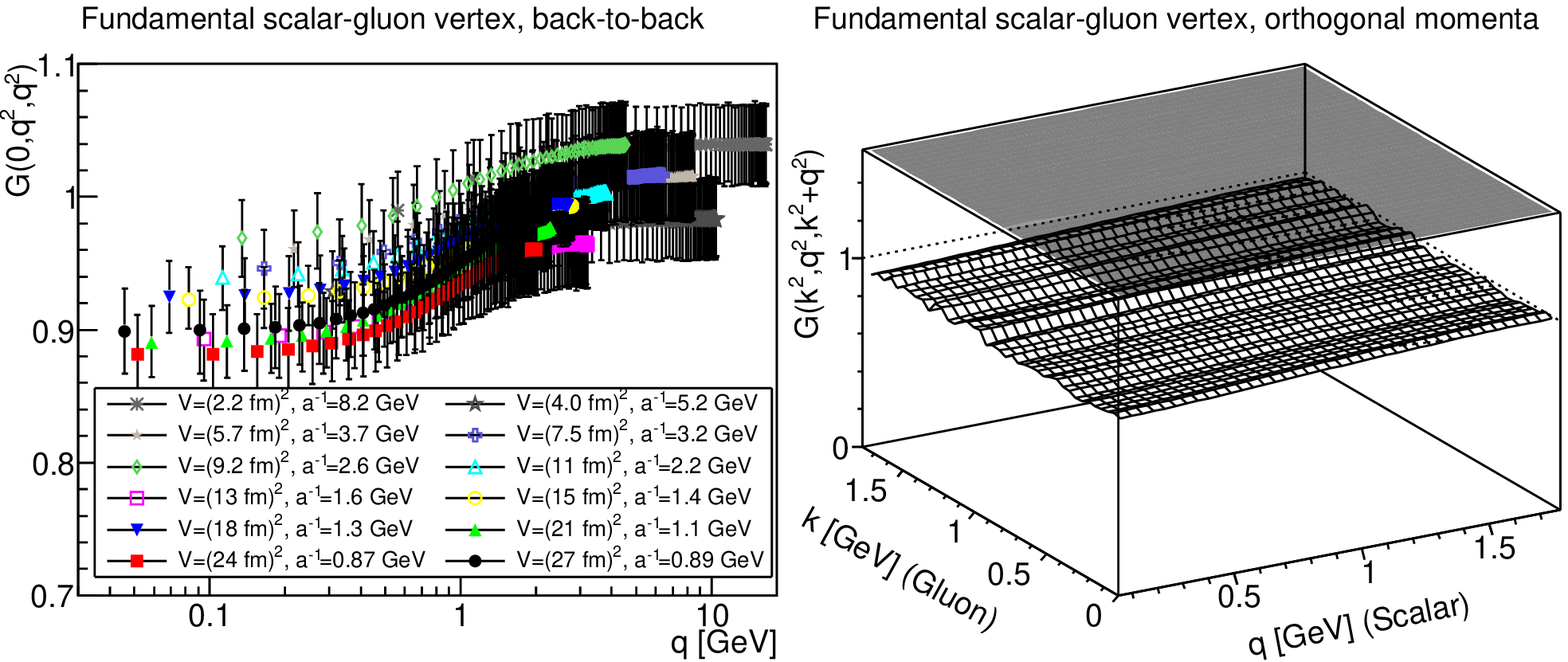}\\
\includegraphics[width=\linewidth]{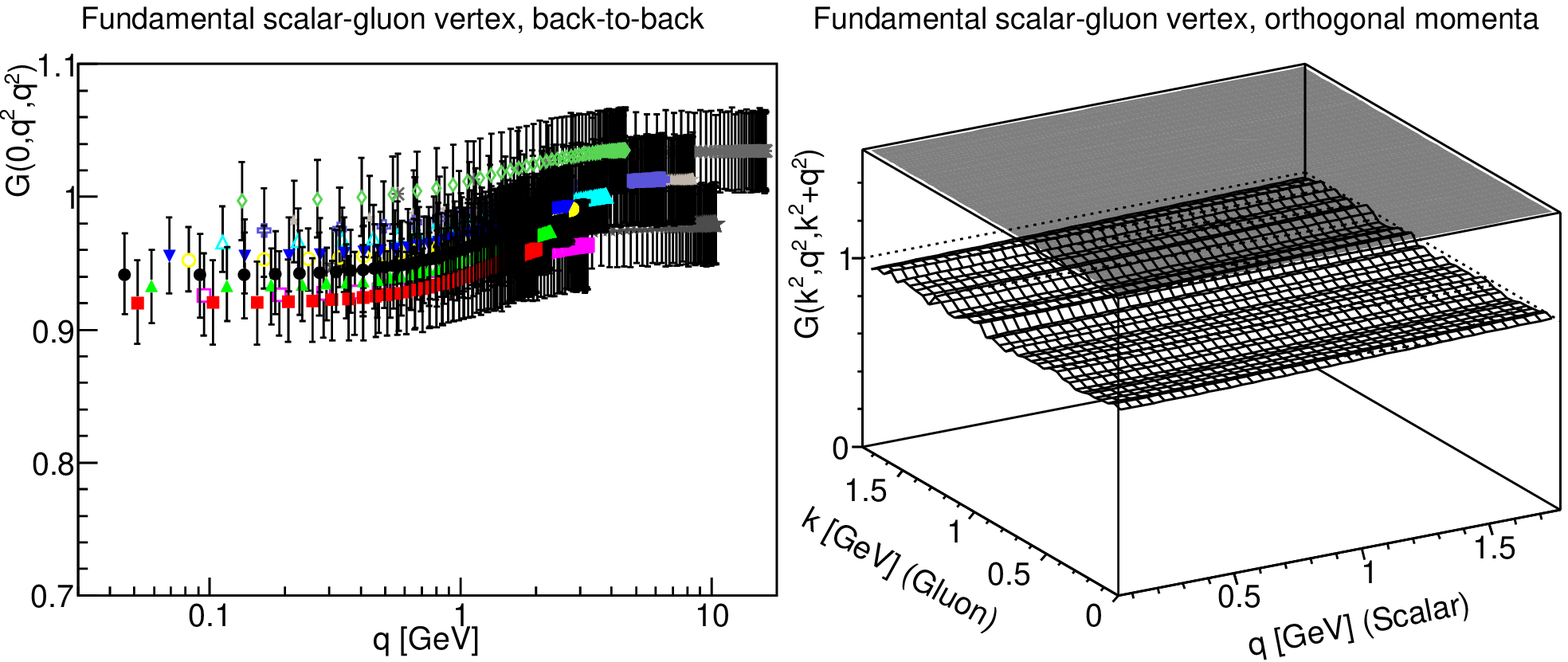}
\caption{\label{fig:2df}The fundamental-scalar gluon vertex in two dimensions. Top panels show $m=0$ GeV and bottom panels $m=1$ GeV. The right panels show the results of the largest volume for all orthogonal momentum configurations, while the left panels show the back-to-back configuration.}
\end{figure}

\begin{figure}
\includegraphics[width=\linewidth]{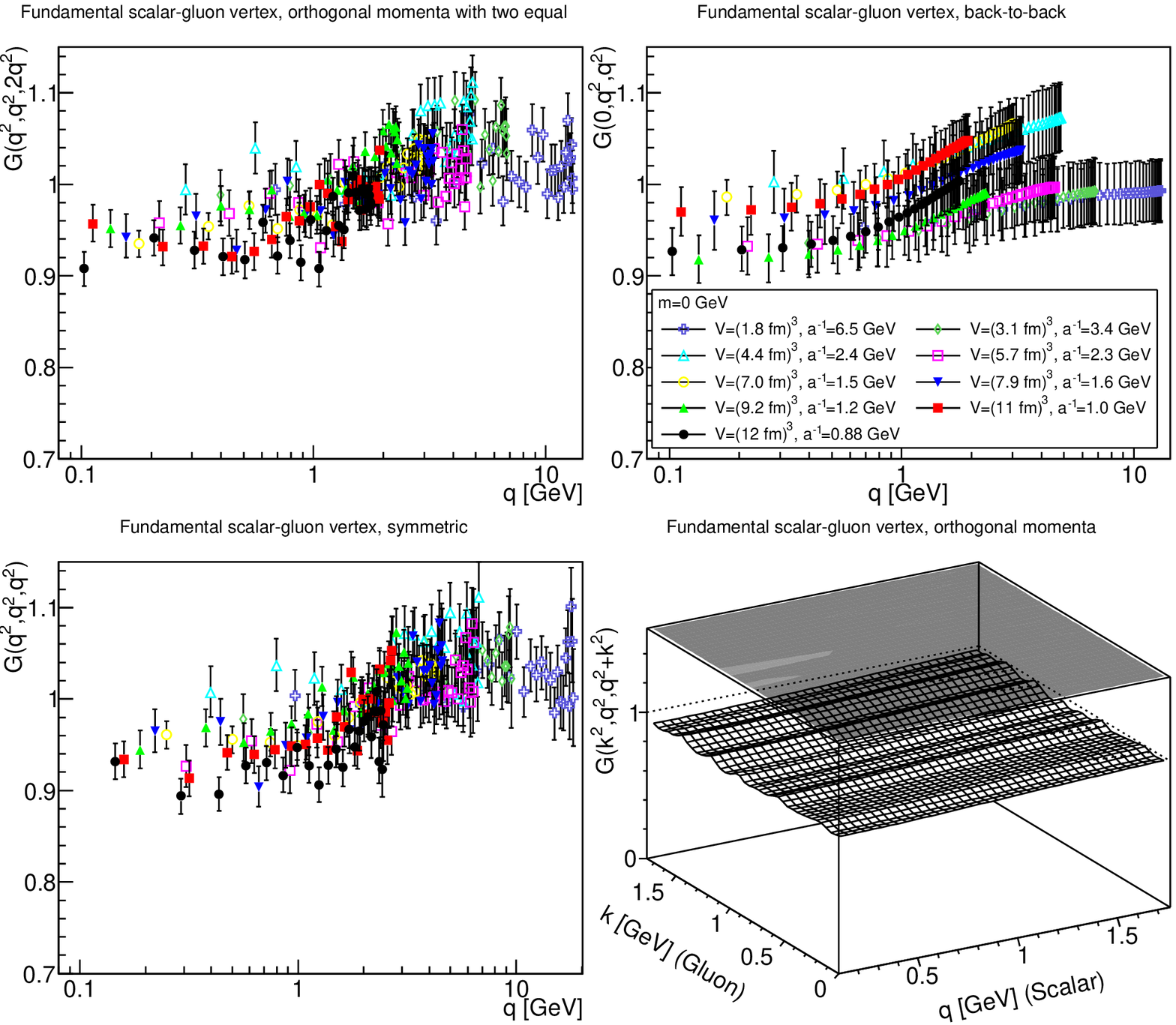}
\caption{\label{fig:3df0}The fundamental-scalar gluon vertex in three dimensions for $m=0$ GeV. The bottom-right panel shows the full dependence in the orthogonal configuration for the largest volume. The top-right panel shows the back-to-back configuration and the top-left panel the orthogonal equal configuration for all volumes at the finest discretization. The lower-left panel shows the symmetric momentum configuration.}
\end{figure}

\begin{figure}
\includegraphics[width=\linewidth]{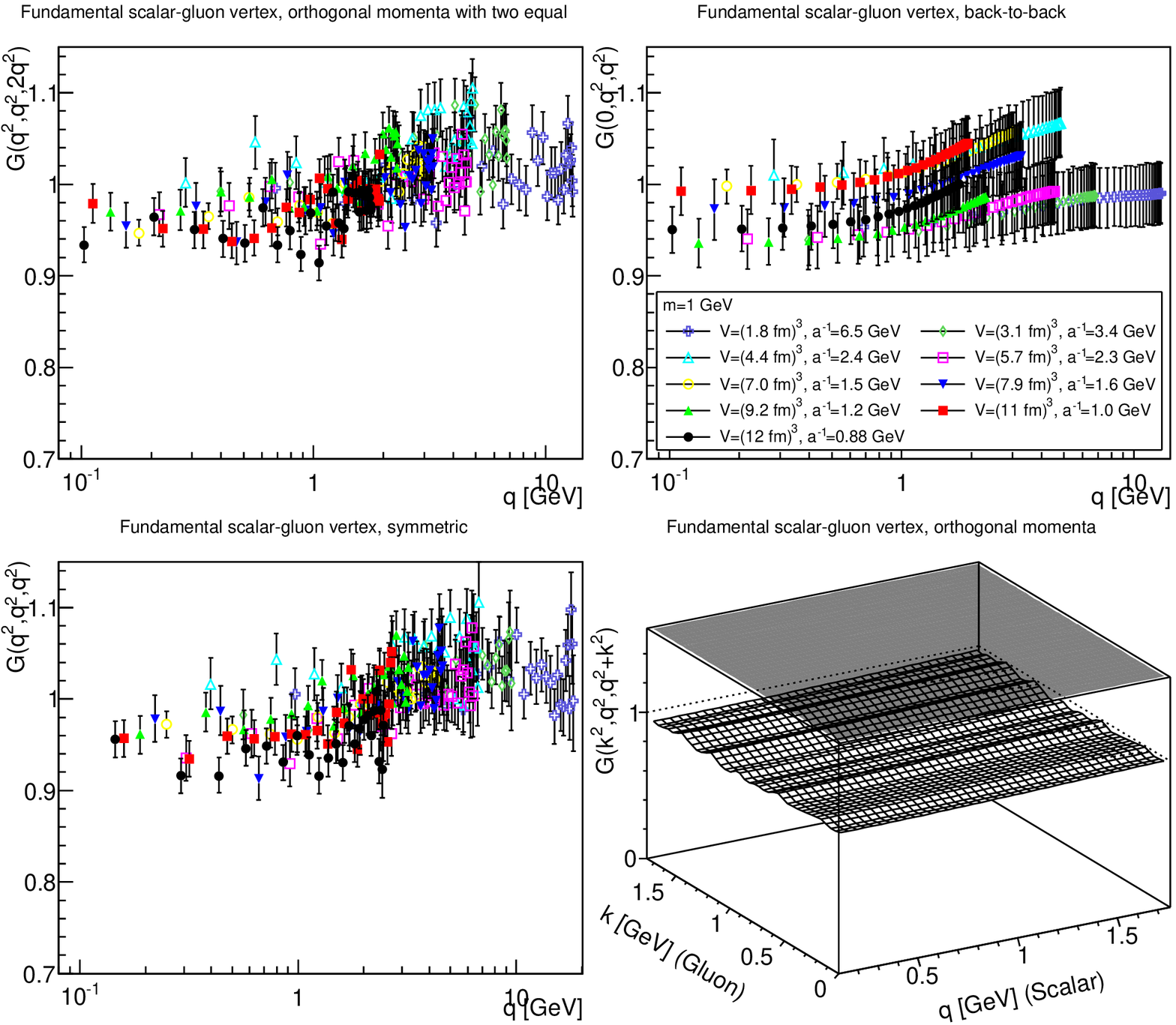}
\caption{\label{fig:3df2}The fundamental-scalar gluon vertex in three dimensions for $m=1$ GeV. The bottom-right panel shows the full dependence in the orthogonal configuration for the largest volume. The top-right panel shows the back-to-back configuration and the top-left panel the orthogonal equal configuration for all volumes at the finest discretization. The lower-left panel shows the symmetric momentum configuration.}
\end{figure}

\begin{figure}
\includegraphics[width=\linewidth]{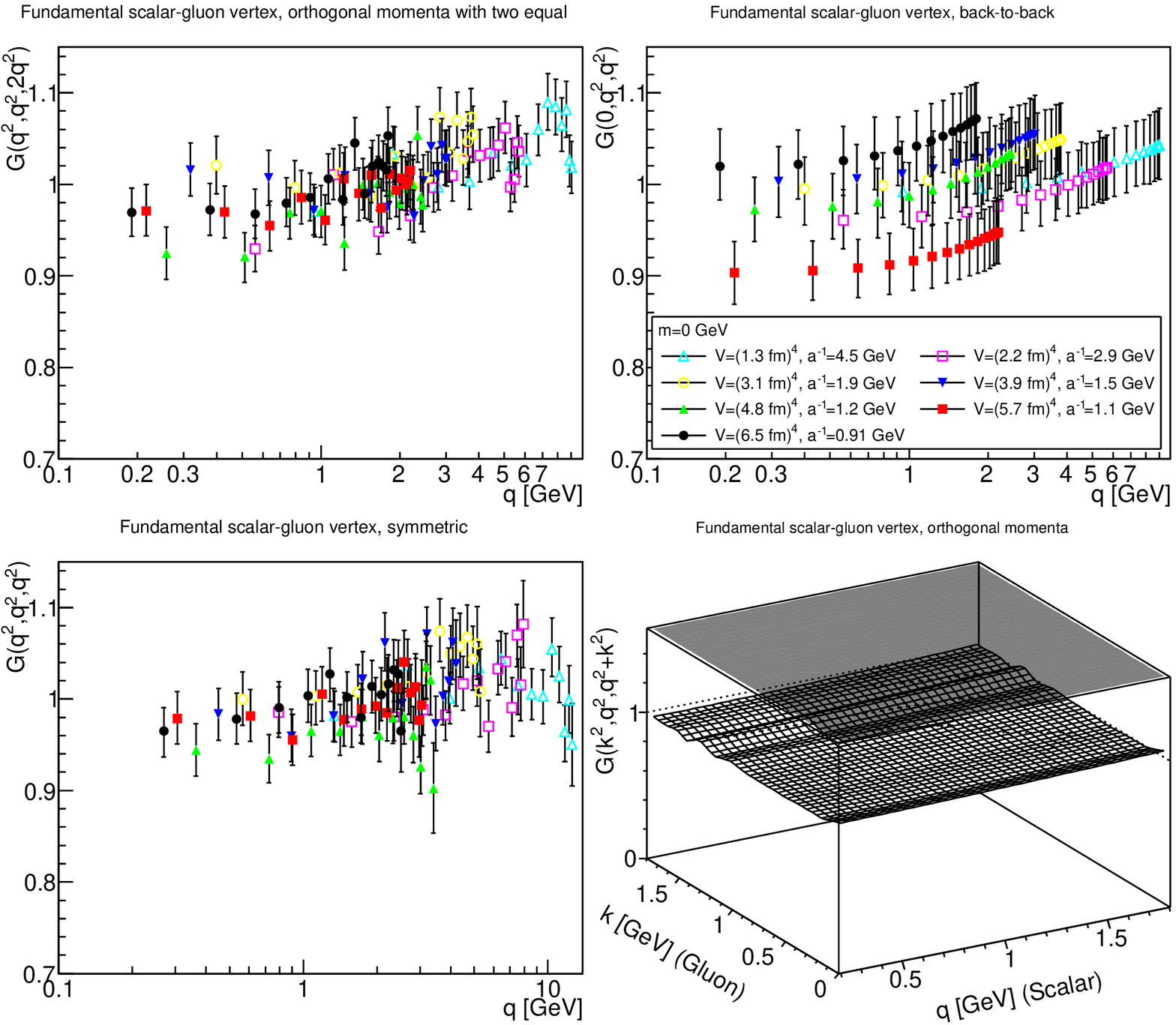}
\caption{\label{fig:4df0}The fundamental-scalar gluon vertex in four dimensions for $m=0$ GeV. The bottom-right panel shows the full dependence in the orthogonal configuration for the largest volume. The top-right panel shows the back-to-back configuration and the top-left panel the orthogonal equal configuration for all volumes at the finest discretization. The lower-left panel shows the symmetric momentum configuration.}
\end{figure}

\begin{figure}
\includegraphics[width=\linewidth]{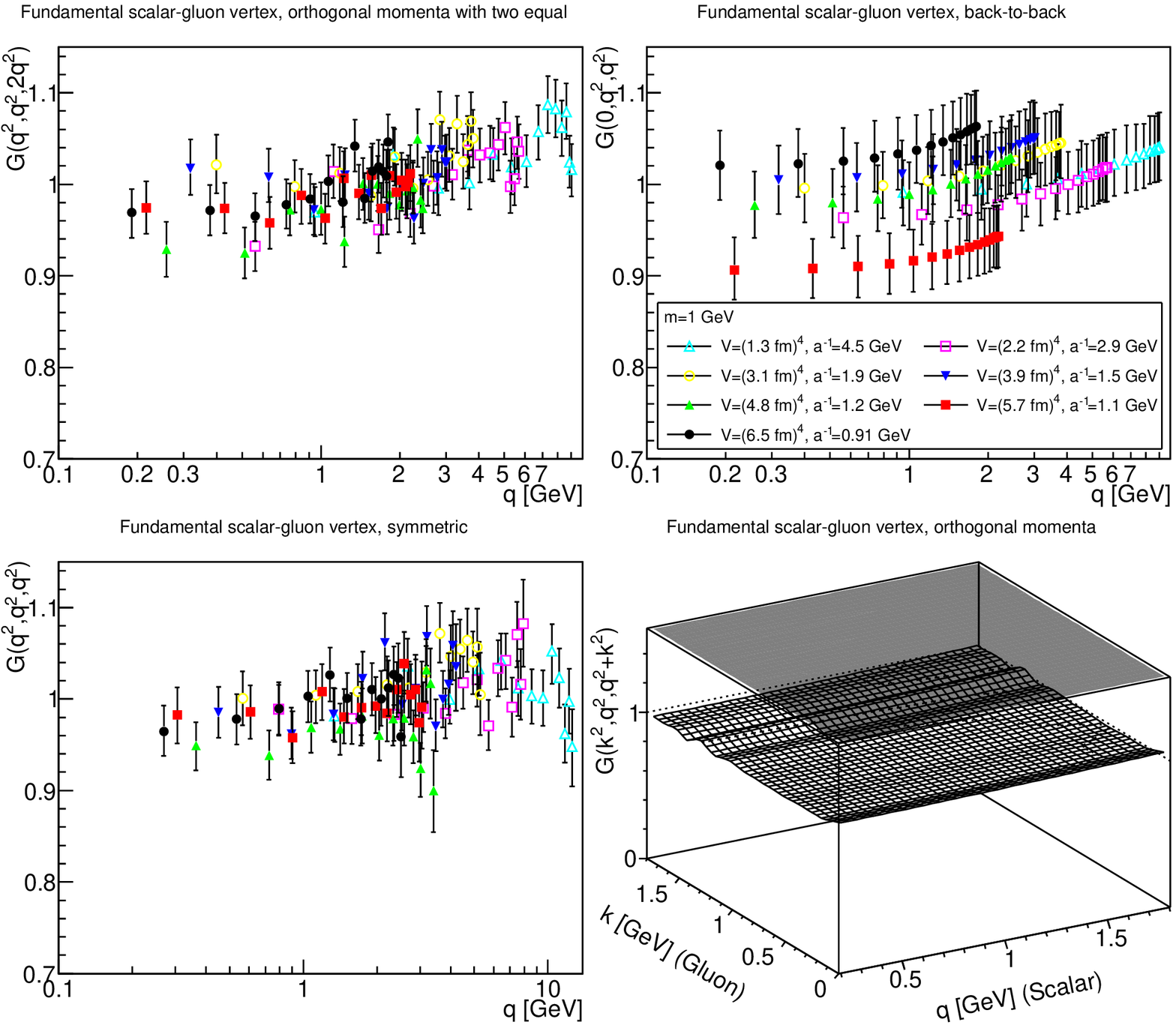}
\caption{\label{fig:4df2}The fundamental-scalar gluon vertex in four dimensions for $m=1$ GeV. The bottom-right panel shows the full dependence in the orthogonal configuration for the largest volume. The top-right panel shows the back-to-back configuration and the top-left panel the orthogonal equal configuration for all volumes at the finest discretization. The lower-left panel shows the symmetric momentum configuration.}
\end{figure}

The results for the fundamental scalar-gluon vertex are shown in figures \ref{fig:2df}-\ref{fig:4df2}. All of them show a qualitatively very similar behavior, interpolating between two (slightly) different values at low momentum and high momentum. The transition between both values occurs in the range between a few hundred MeV and 2 GeV. This also occurs at zero gluon momentum, indicating that the scalar momentum is the relevant quantity.

The ratio of the ultraviolet to the infrared constant depends on the mass, and is larger the smaller the mass. However, with increasing dimensionality the ratio becomes less and less dependent on the mass, until in four dimensions the results at $m=0$ GeV and $m=1$ GeV cannot be distinguished by the naked eye, and are, in fact, smaller than the statistical error.

The at first sight different behavior between zero gluon momentum and non-zero gluon momentum, especially the apparent correlation of the form factor at zero gluon momentum, has a simple\footnote{Of course, there may be other effects relevant as well. Without having all momentum configurations available, this cannot be excluded.} explanation. The scalar field is quenched, and the inversion of the Laplacian occurs for all momenta on the same field configurations. At zero gluon momentum this is leads to highly correlated results at different momentum, because there always the same Fourier mode of the gluon field enters, and thus only the ordinary Laplacian in \pref{covf} changes from momentum to momentum. This is not the case for any other momentum configuration, as there always different Fourier modes of the gluon fields enter, leading to less correlated results. As a consequence, this leads to the apparent shifts in the back-to-back configuration. Because the fluctuations are correlated, any fluctuation above or below the renormalized value will be the same for all momenta. Thus, the non-zero gluon momentum cases give a much better idea of the actual scattering of the data around the true form factor, especially as the renormalization is performed according to \pref{rcond}. If the renormalization would have been performed at zero gluon momentum, no correlated shifts would be present. Note that this also applies to the adjoint case in section \ref{ss:adjoint}.

These results are consistent with the dynamical case, where in the QCD-like case an essentially constant gluon-scalar form factor is found \cite{Maas:2013aia}. However, the statistical errors in \cite{Maas:2013aia} were too large to detect the slight drop when going form the ultraviolet to the infrared. Still, this indicates that unquenching effects are likely small for this vertex. Note that also in the Brout-Englert-Higgs region of this theory the scalar-gluon vertex is not substantially modified to the present quenched case \cite{Maas:2013aia}.

Thus, in total the fundamental scalar-gluon vertex does not show any significant deviation from the tree-level behavior. In fact, it is even less affected than the ghost-gluon vertex, which showed hitherto the smallest deviations from tree-level \cite{Cucchieri:2006tf,Cucchieri:2008qm,Maas:2007uv,Cucchieri:2004sq,Sternbeck:2005re,Sternbeck:2012mf}.

\subsection{Adjoint case}\label{ss:adjoint}

\begin{figure}
\includegraphics[width=\linewidth]{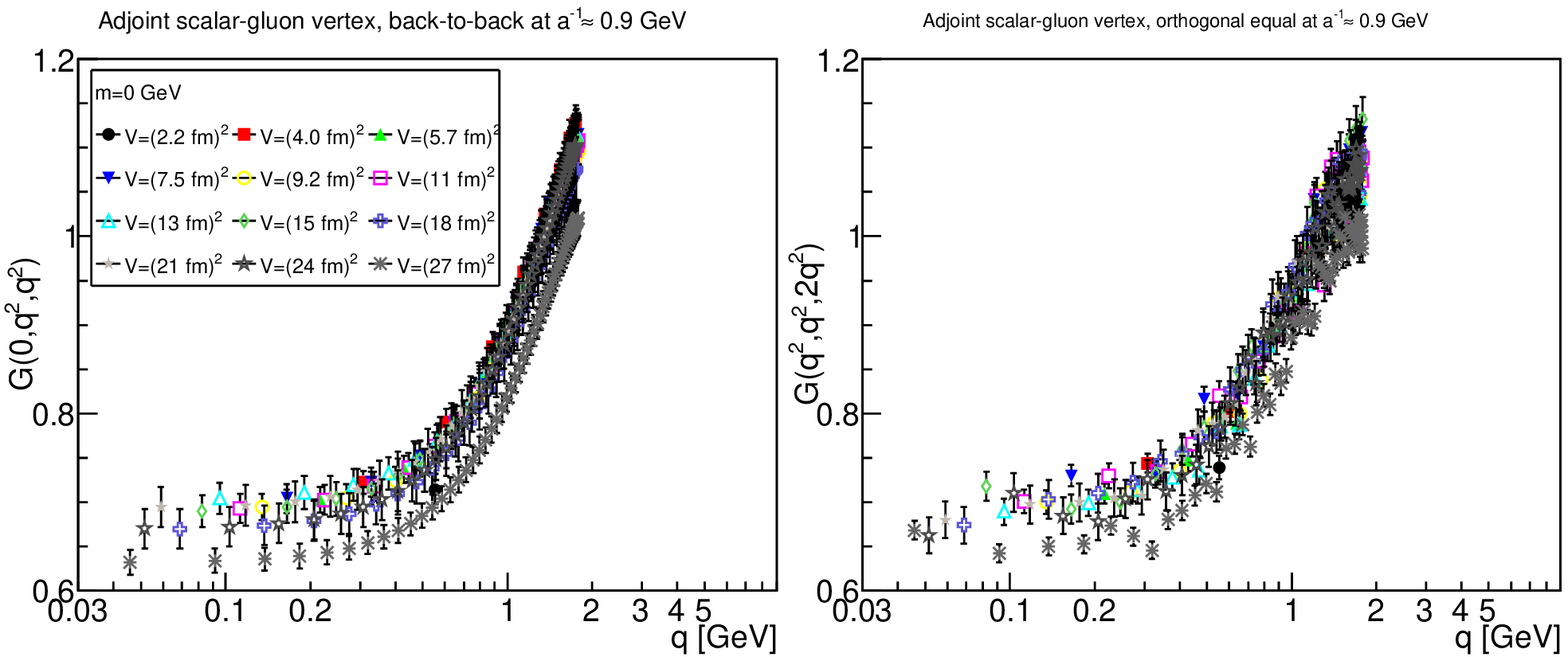}
\includegraphics[width=\linewidth]{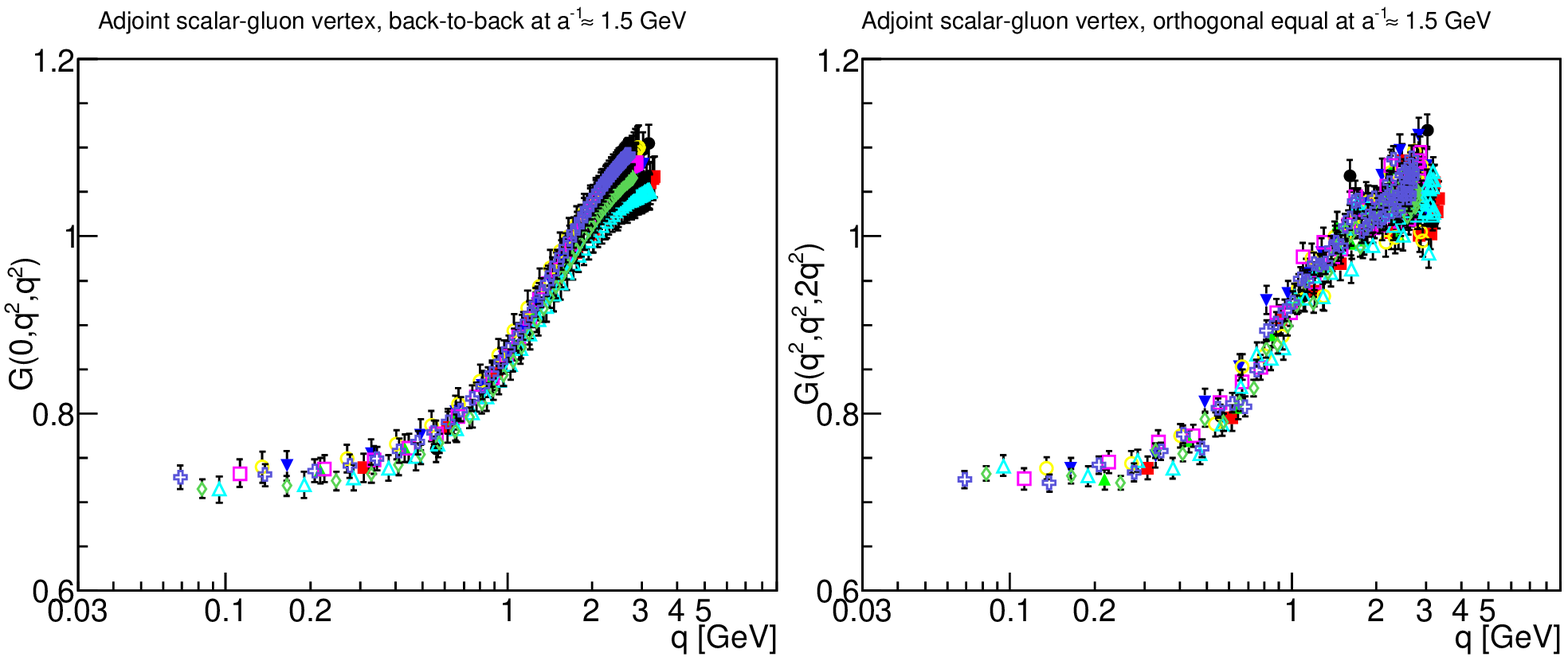}
\includegraphics[width=\linewidth]{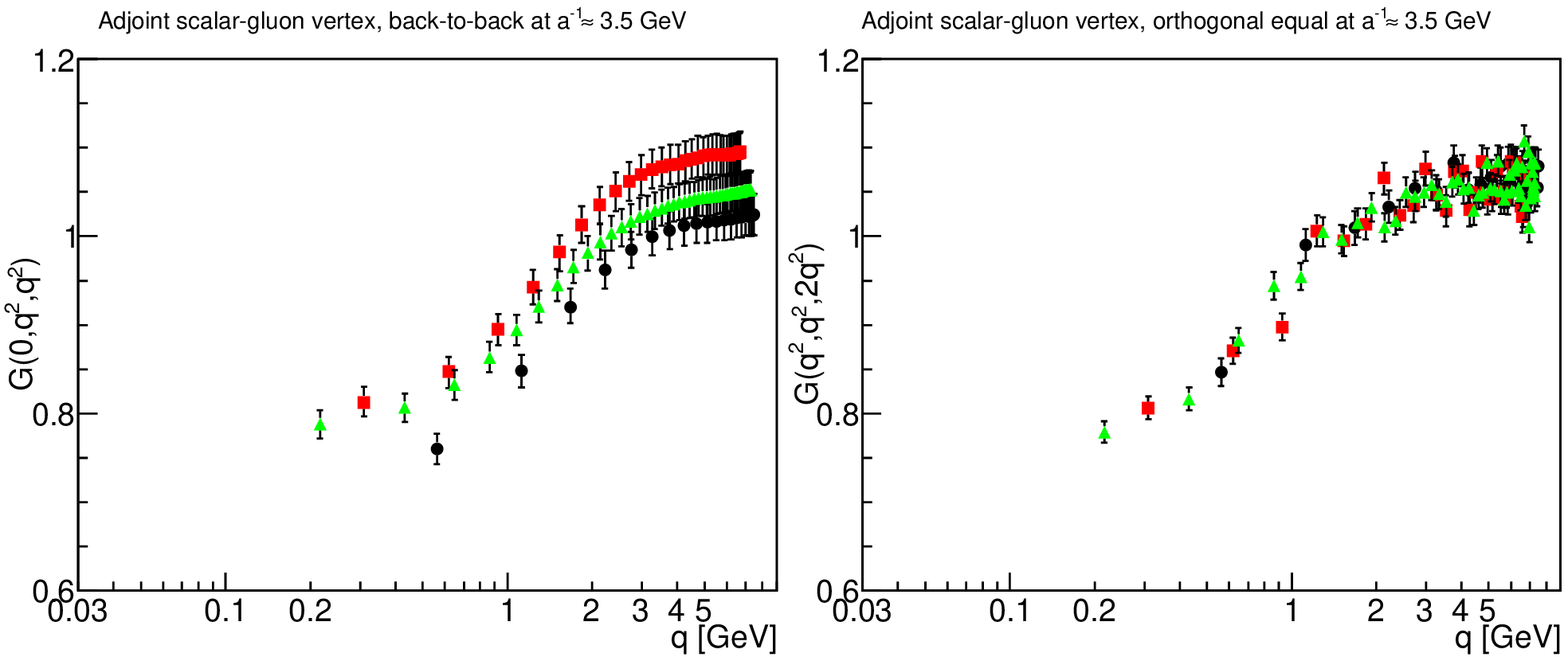}
\caption{\label{fig:2da09}The adjoint scalar-gluon vertex in two dimensions for $m=0$ GeV and fixed discretization of $a^{-1}\approx 0.9$ GeV (top panels) $^a{-1}\approx 1.5$ GeV (middle panels) and $a^{-1}\approx 3.5$ GeV (bottom panels) for the back-to-back (left panels) and orthogonal equal (right panels) momentum configurations.}
\end{figure}

\begin{figure}
\includegraphics[width=\linewidth]{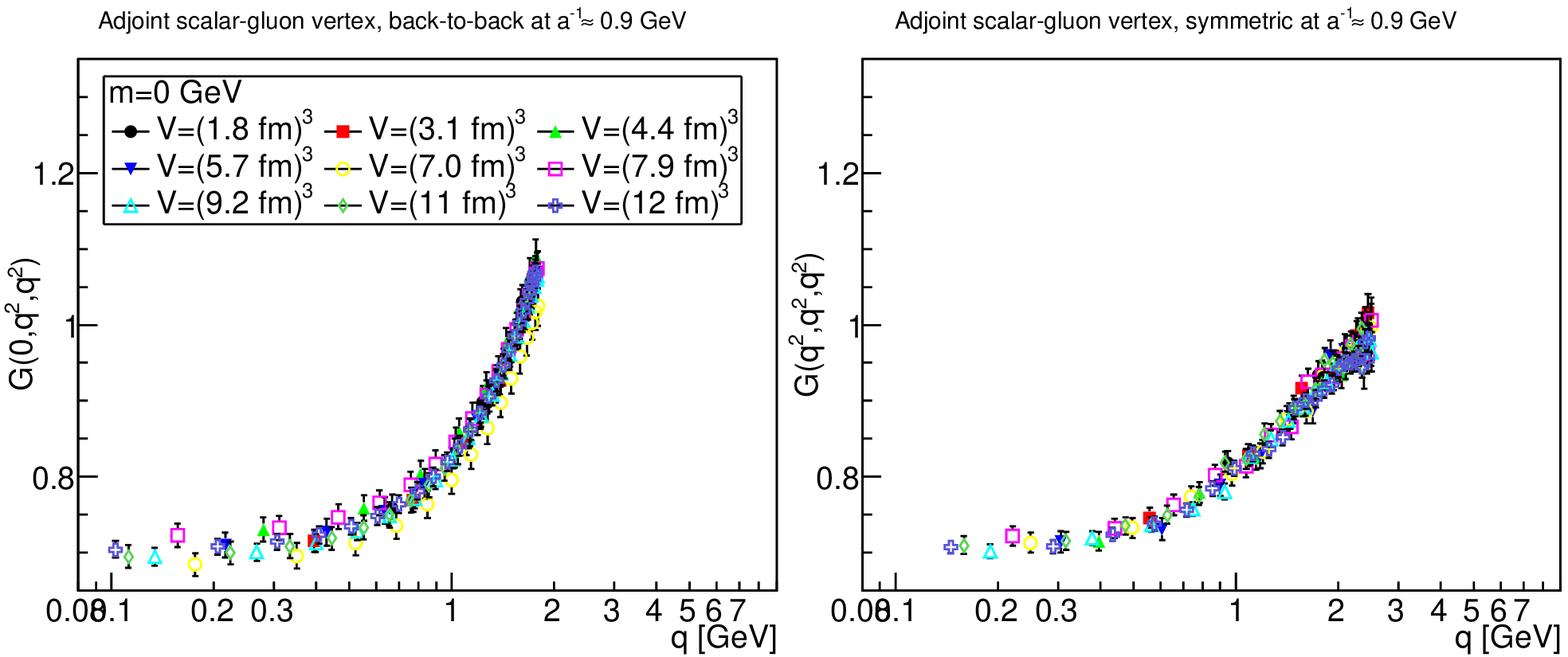}
\includegraphics[width=\linewidth]{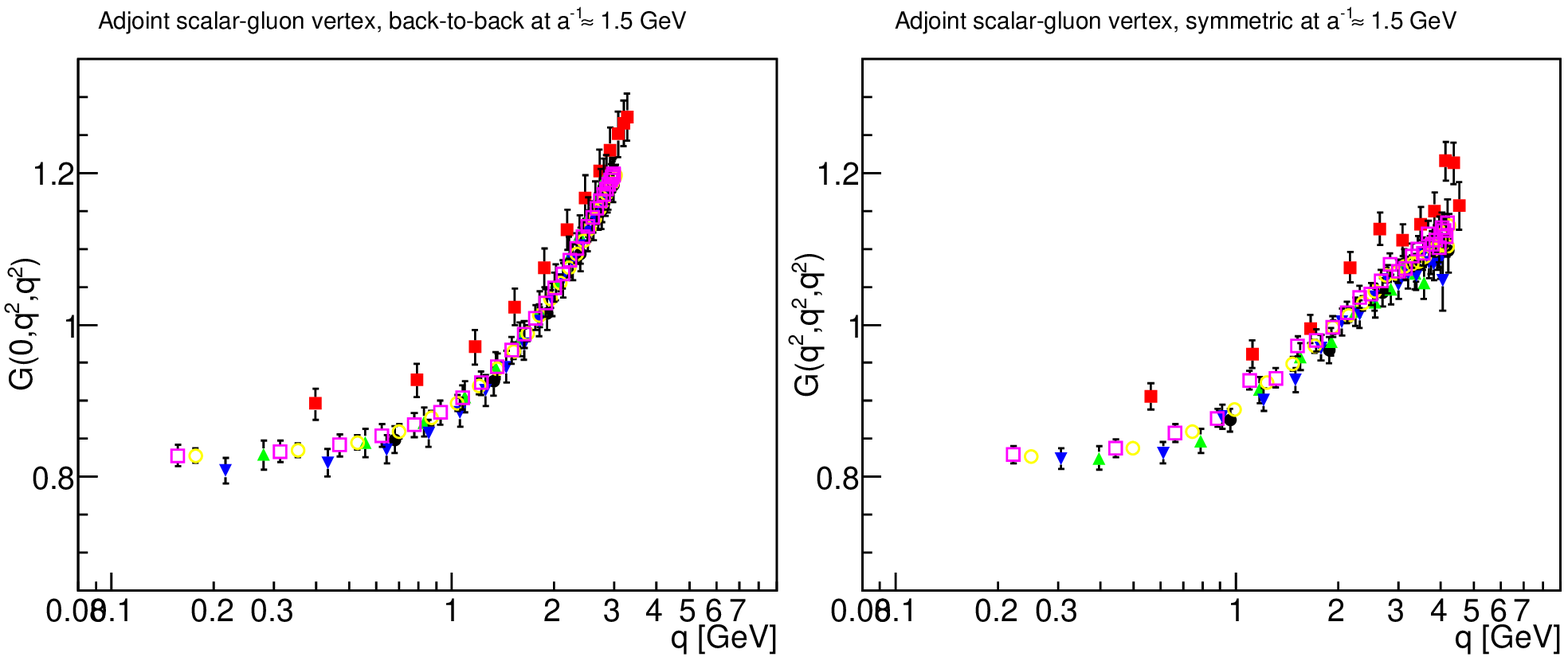}
\includegraphics[width=\linewidth]{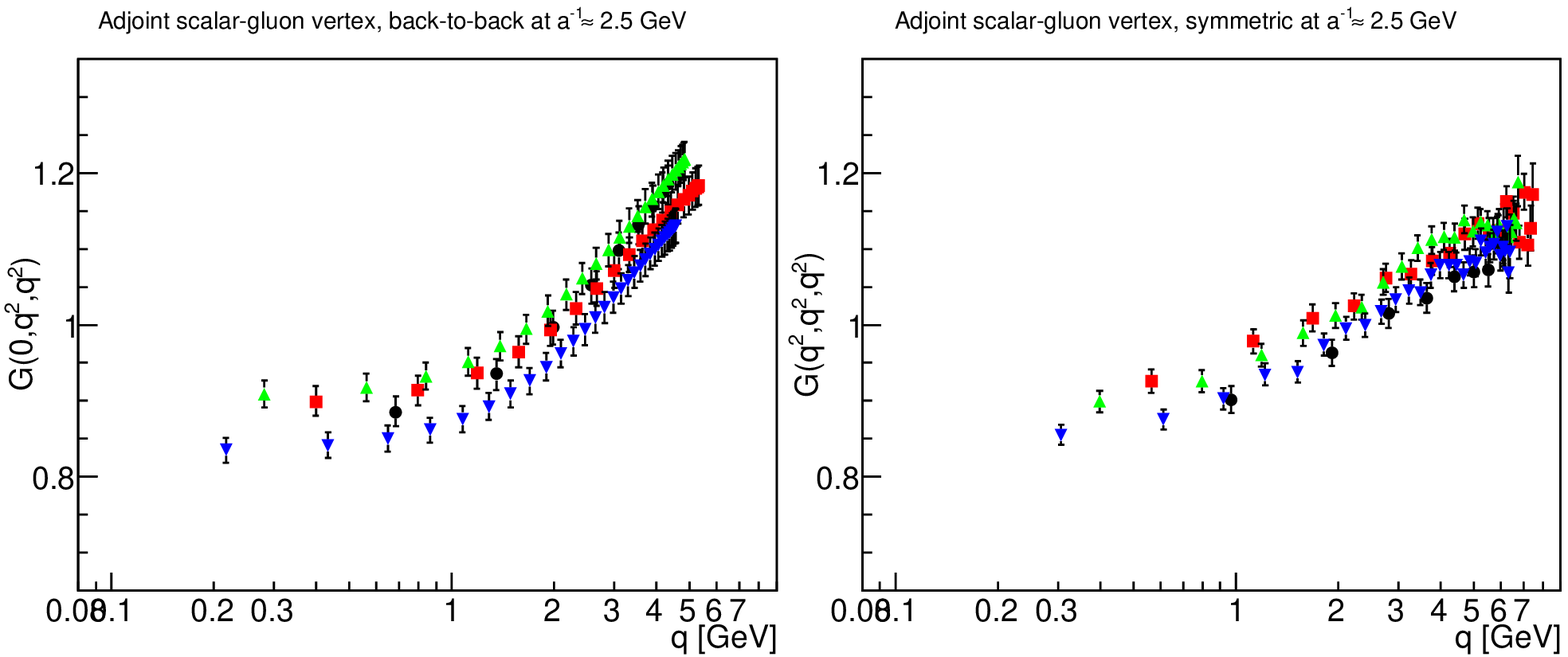}
\caption{\label{fig:3da09}The adjoint scalar-gluon vertex in three dimensions for $m=0$ GeV and fixed discretization of $a^{-1}\approx 0.9$ GeV (top panels) $a^{-1}\approx 1.5$ GeV (middle panels) and $a^{-1}\approx 2.5$ GeV (bottom panels) for the back-to-back (left panels) and symmetric (right panels) momentum configurations.}
\end{figure}

\begin{figure}
\includegraphics[width=\linewidth]{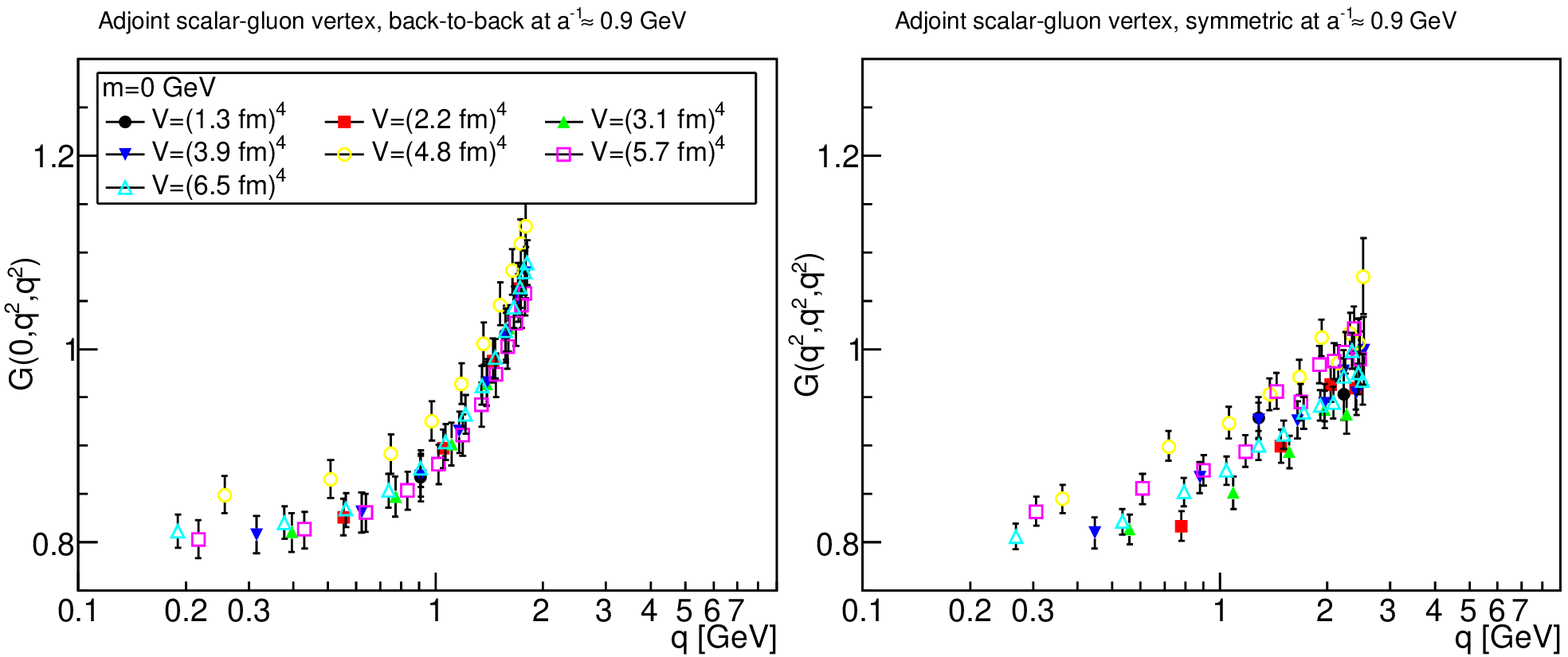}
\includegraphics[width=\linewidth]{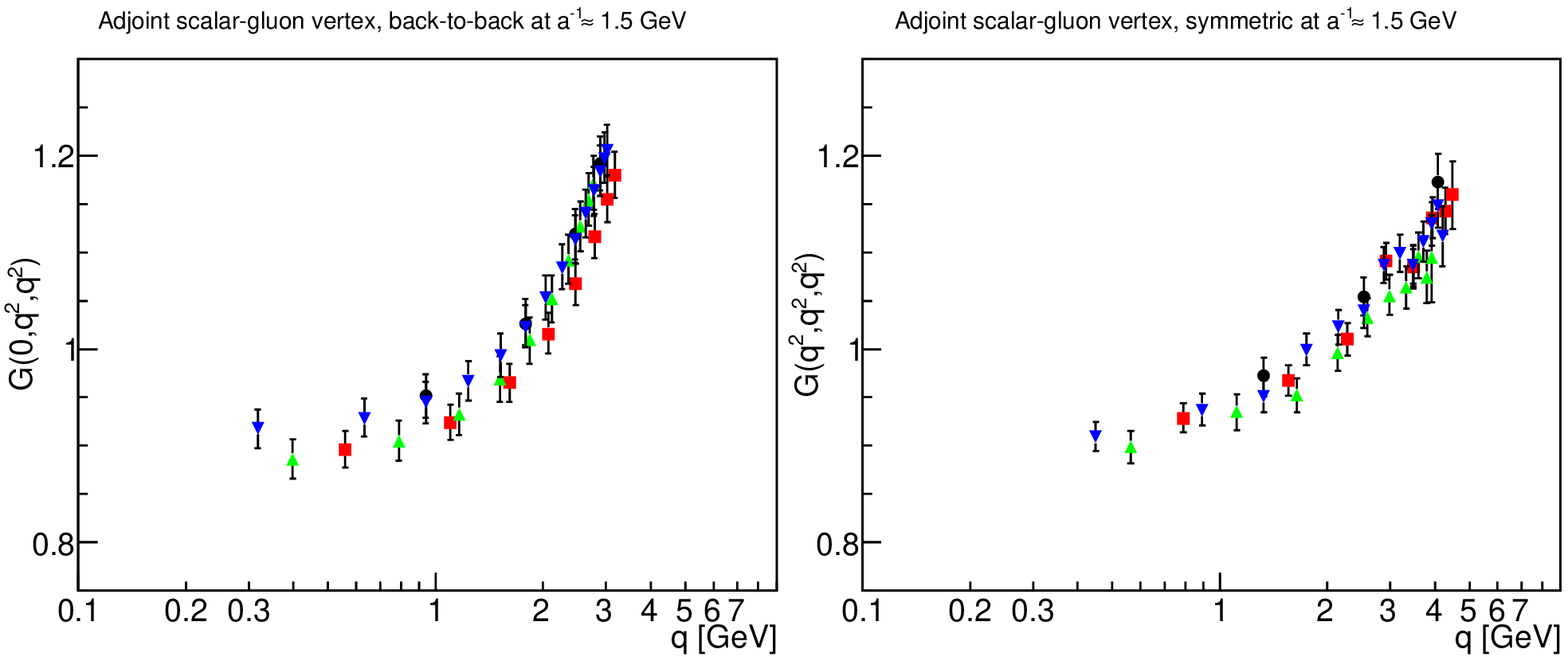}
\includegraphics[width=\linewidth]{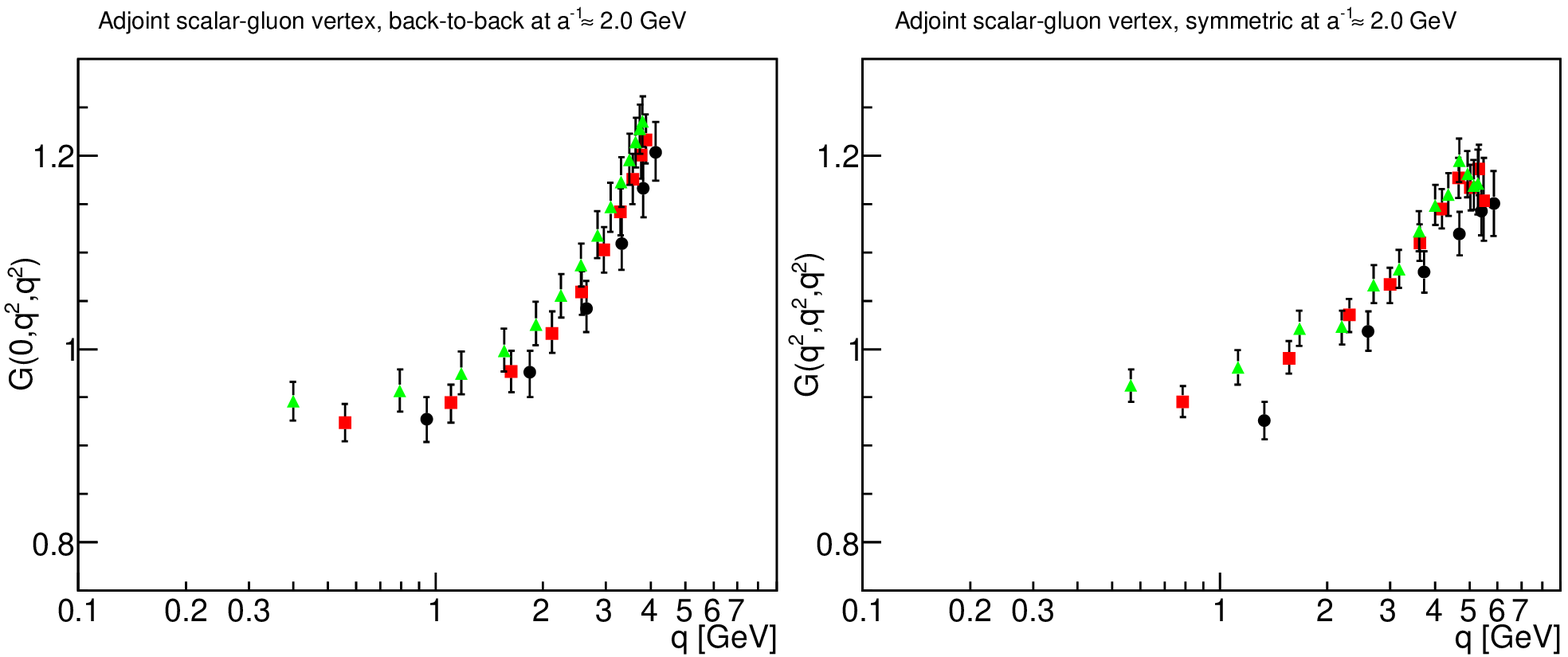}
\caption{\label{fig:4da09}The adjoint scalar-gluon vertex in four dimensions for $m=0$ GeV and fixed discretization of $a^{-1}\approx 0.9$ GeV (top panels), $a^{-1}\approx 1.5$ GeV (middle panels) and $a^{-1}\approx 2.0$ GeV (bottom panels) for the back-to-back (left panels) and symmetric (right panels) momentum configurations.}
\end{figure}

As noted already in section \ref{s:ren}, the adjoint vertex is, as the adjoint propagator \cite{Maas:2018sqz}, much stronger affected by discretization artifacts. To better assess them, the vertex is shown at various fixed discretizations for two, three, and four dimensions in figures \ref{fig:2da09}-\ref{fig:4da09}, respectively. For this purpose, the situation at $m=0$ is considered exclusively, as its largest deviation from tree-level enhances all effects most. Likewise, the angular dependence of the discretization artifacts are small, and thus it is sufficient to look at two particular momentum settings.

The results in two dimensions in figure \ref{fig:2da09} show very little volume dependence at fixed discretization. Generically, there is a transition between two constant regimes between about 300 MeV to about 3 GeV. This is a relatively slow transition. The appearance of the high-momentum constant behavior is only clearly visible when correspondingly large momenta above 3 GeV can be reached. The drop towards the infrared depends on the discretization, getting slightly smaller on finer lattices, but amounts to at most about 40\%. Thus, by and large, the scalar-gluon vertex is only weakly deviating from tree-level, but still substantially more so than the fundamental one. 

The situation in three dimensions, shown in figure \ref{fig:3da09}, is quite similar. There is no pronounced volume-dependence at fixed discretization. However, because only a smaller separation of scales can be achieved in the more expensive three-dimensional setting the reaching of the ultraviolet plateau becomes only visible at the finest discretization in the symmetric momentum regime. The drop between both plateaus is slightly large than in two dimensions, about 50\%, and the ultraviolet plateau seems to form only around 4 GeV, but the infrared one already at 400 MeV.

The picture repeats itself in four dimensions, shown in figure \ref{fig:4da09}, without a strong volume dependence at fixed discretization. The ultraviolet plateau is once more reached later, now at about 5 GeV, and the infrared at about 500 MeV. However, because again the maximal scale separation is smaller, this now only indicates itself. The drop itself is of the same order as in three dimensions.

In total, volume effects therefore tend to increase the infrared value slightly, while discretization leads to a flattening at large momenta. The transition region between the asymptotic regimes is about one order of magnitude in momentum. The transition starts at the same typical scale already observed for the properties of the propagators, a few hundred MeV \cite{Maas:2016edk,Maas:2018sqz}. Note that the actual infrared value itself is not significant. By altering the renormalization prescription it would have been possible to fix the infrared value for all lattice setups to the same value, e.\ g.\ one. Then all effects would have become ultraviolet effects in the form of a flattening towards an asymptotic value from above at finer and finer discretization, with virtually no volume dependence.

\begin{figure}
\includegraphics[width=\linewidth]{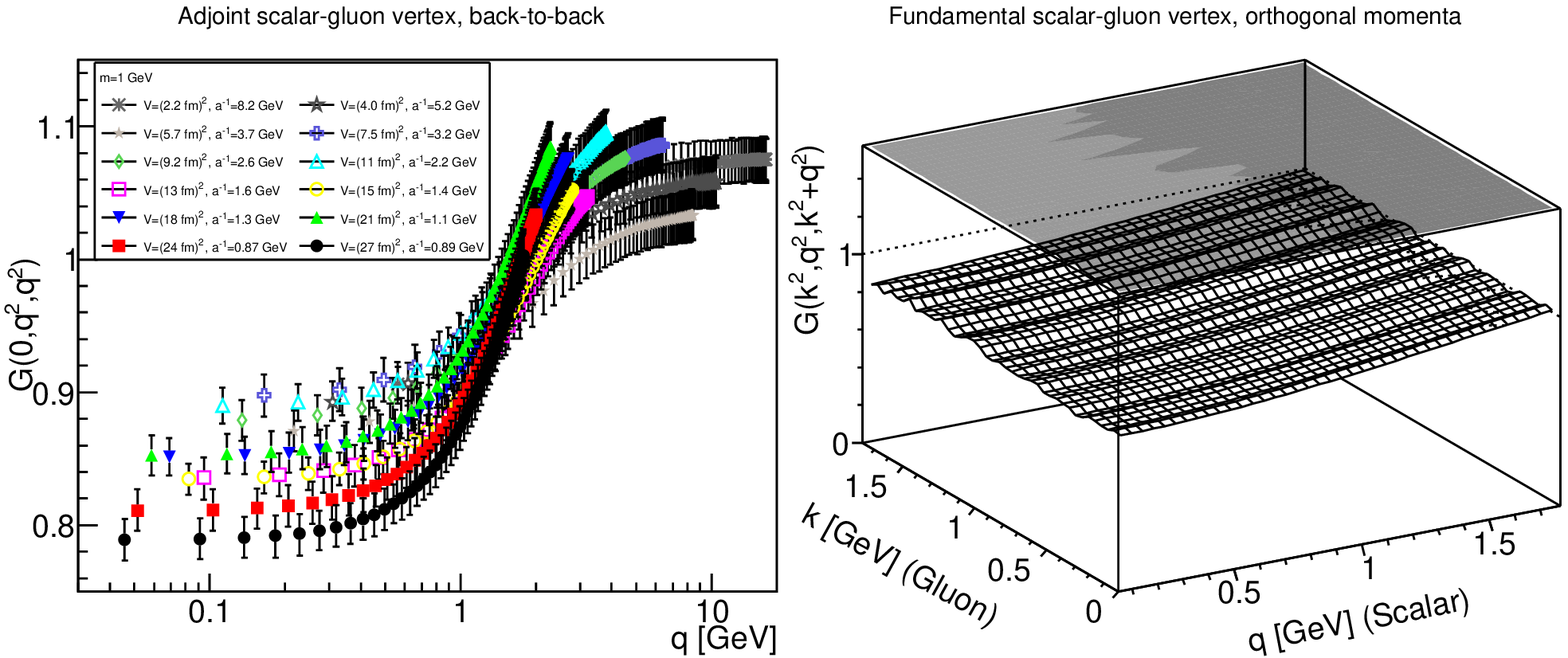}
\caption{\label{fig:2da}The adjoint-scalar gluon vertex in two dimensions for $m=1$ GeV. The right panel shows the results of the largest volume for all orthogonal momentum configurations, while the left panel shows the back-to-back configuration.}
\end{figure}

\begin{figure}
\includegraphics[width=\linewidth]{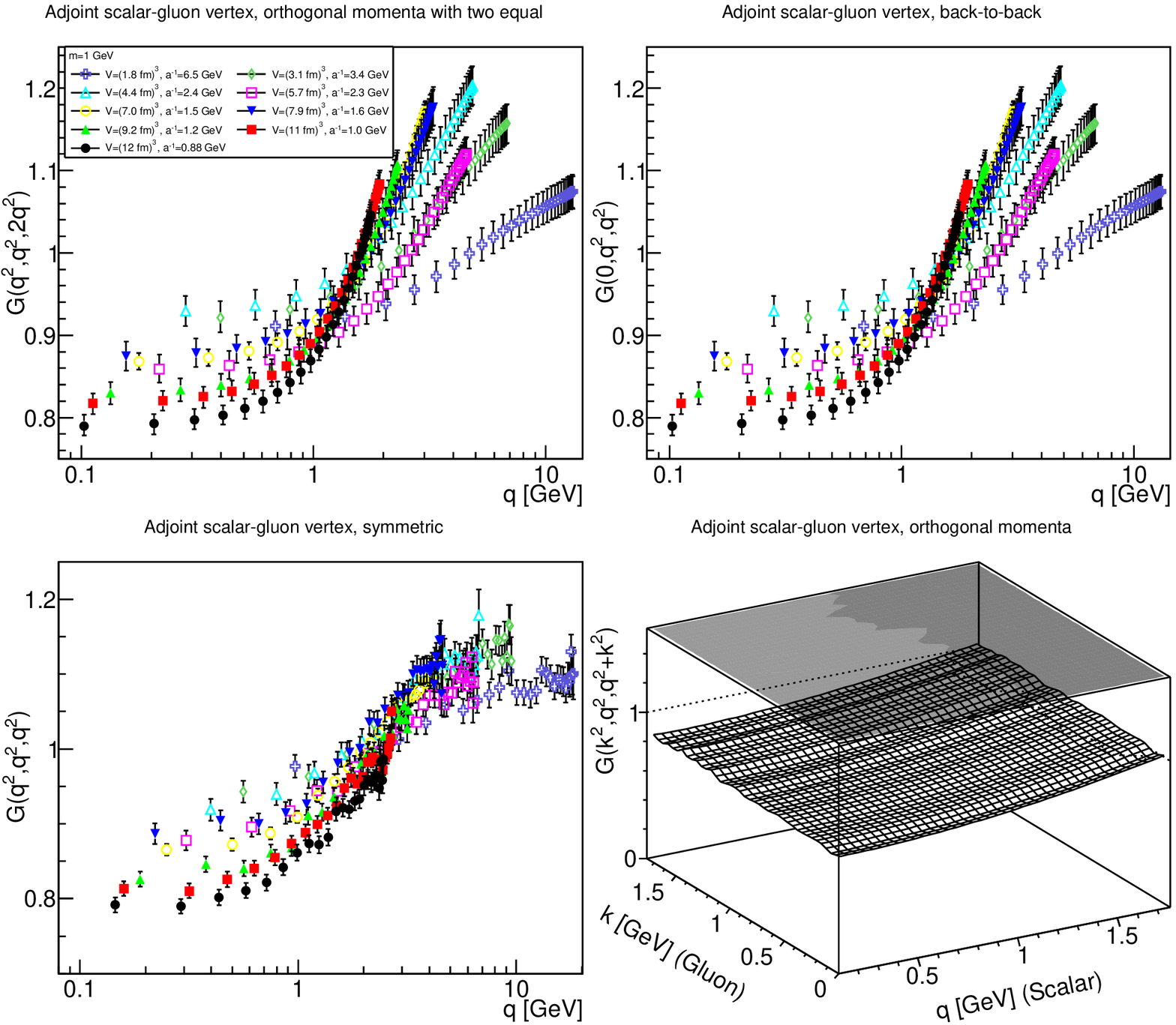}
\caption{\label{fig:3da}The adjoint-scalar gluon vertex in three dimensions for $m=1$ GeV. The bottom-right panel shows the full dependence in the orthogonal configuration for the largest volume. The top-right panel shows the back-to-back configuration and the top-left panel the orthogonal equal configuration for all volumes at the finest discretization. The lower-left panel shows the symmetric momentum configuration.}
\end{figure}

\begin{figure}
\includegraphics[width=\linewidth]{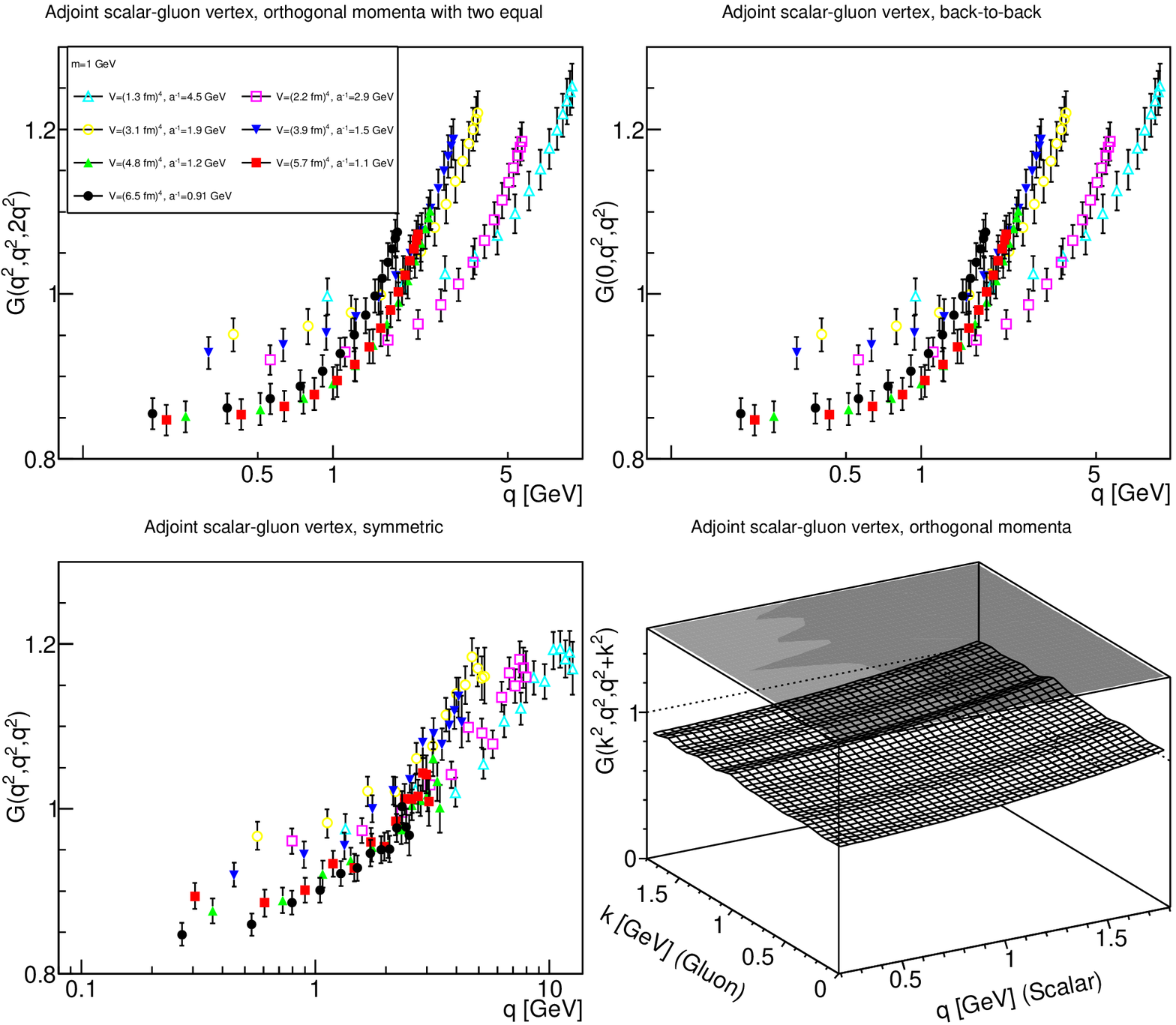}
\caption{\label{fig:4da}The adjoint-scalar gluon vertex in four dimensions for $m=1$ GeV. The bottom-right panel shows the full dependence in the orthogonal configuration for the largest volume. The top-right panel shows the back-to-back configuration and the top-left panel the orthogonal equal configuration for all volumes at the finest discretization. The lower-left panel shows the symmetric momentum configuration.}
\end{figure}

Keeping this in mind, figures \ref{fig:2da}-\ref{fig:4da} shows the adjoint form factor for all momentum configurations at the finest discretizations for all volumes, but now for $m=1$ GeV to study also the mass dependence.

The two-dimensional case in figure \ref{fig:2da} shows the general behavior quite nicely. The form factor gets squeezed with decreasing lattice spacing, keeping its value at the largest momenta, more or less, and increases the infrared plateau. Studying the full momentum dependence shows that the transition is driven essentially by the scalar momentum, while the result is relatively independent of the gluon momentum. In comparison to the zero mass case in figure \ref{fig:2da09} the drop gets smaller, which is mainly pushed into the infrared as a consequence of the renormalization condition. As noted above, this continues for increasing mass, until for the 10 GeV case the form factor is essentially flat. Thus, the mass dependence is merely that the lighter the mass the larger the drop, though the drop remains limited. There is no visible sign of any singular behavior, neither at the actual momenta measured nor as tendency towards the thermodynamic limit.

The situation is quite similar in three dimensions, shown in figure \ref{fig:3da}. Interestingly, the lattice artifacts are stronger for the orthogonal configuration than for the symmetric configuration, even at finite gluon momentum. Still, the qualitative behavior is as in two dimensions. At sufficiently fine discretization there is again a drop, which is slightly smaller than in the zero-mass case. Also, the scalar momentum seems to be the driver of this behavior, as it appears in the orthogonal configuration quite independently of the gluon momentum.

Unsurprisingly, the same pattern emerges in four dimensions as in lower dimensions, as shown in figure \ref{fig:4da}, just with a little bit different scales.

\section{Conclusions}\label{s:con}

The presented investigation of the vertices shows, as for the propagators \cite{Maas:2016edk,Maas:2018sqz}, unfortunately no obvious sign of how confinement in the Wilson sense acts differently on the fundamental and adjoint charges. In fact, the form factors do not show any substantial deviations from the tree-level at all, similarly to the ghost-gluon vertex \cite{Cucchieri:2006tf,Cucchieri:2008qm,Maas:2007uv,Cucchieri:2004sq,Sternbeck:2005re,Sternbeck:2012mf,Zafeiropoulos:2019flq,Huber:2018ned}, but very different from the three-gluon vertex \cite{Cucchieri:2008qm,Maas:2007uv,Sternbeck:2017ntv,Huber:2018ned,Boucaud:2017obn,Binosi:2017rwj}.

Especially, no singularities are observed. However, singularities are mainly expected within a so-called scaling behavior of the propagators \cite{Fister:2010yw,Zwanziger:2010iz,Alkofer:2006gz,Alkofer:2008tt,Schwenzer:2008vt}, while a finite behavior is associated with a screening (or decoupling) behavior \cite{Fister:2010yw}. The propagator in the presently used gauge are of the latter type \cite{Maas:2011se}, and for this infrared finite vertices are expected \cite{Fister:2010yw}. Thus, one could interpret this as a consistent behavior. However, in two dimensions a scaling behavior prevails \cite{Maas:2011se}, and still no singularities are observed. Such a combination has also been speculated about \cite{Fister:2010yw}, and thus seems to be also realizable.

The results are also substantially different from those of the quark-gluon vertex, where stronger deviations from the tree-level case are seen \cite{Alkofer:2006gz,Alkofer:2008tt,Kizilersu:2006et,Sternbeck:2017ntv,Binosi:2016wcx,Huber:2018ned,Eichmann:2014xya}. If one assumes that confinement for scalars and fermions works in the same way, it should manifest itself in the same way in their vertices. Any differences in the vertices should then be due to differences between fermions and scalars, which is mainly chiral symmetry. This suggests the speculation that the deviations of the quark-gluon vertex from tree-level are more related to chiral symmetry breaking than confinement, though this is not a necessary consequence.

Taking a different perspective on this result, the agreement between two and higher dimensions could also be interpreted differently. After all, confinement in two dimensions cannot be due to dynamics, as the gauge sector is non-dynamical. Any results in the quenched theory are therefore necessarily due to the gauge structure only. Especially, confinement cannot arise due to a dynamical effect, but must be structural in origin. As the interactions with the scalars in two and higher dimensions is the same, this suggests that to be also true in higher dimensions, provided the vertices even without singularities encode how confinement operates. Such a structural origin of confinement would concur with equating confinement entirely with gauge invariance when taking the Gribov-Singer ambiguity fully into account \cite{Lavelle:1995ty,Maas:2017wzi}.

Considering the form factor as an entity independent of this question, it shows, however, a behavior quite different from what one would expect for a physical particle \cite{Pacetti:2015iqa}. First, for a physical, charged particle it is expected that, when probed by a current coupling to this charge, it has an increasing form factor, not a decreasing one as the one seen here. This is also a characteristic feature shown by many other gauge-dependent particles \cite{Maas:2011se,Maas:2018ska}, especially gluons. Another feature is that the derivative of the form factor at zero momentum can be interpreted as a radius \cite{Pacetti:2015iqa}. For the scalar-gluon vertex here the intercept is essentially constant. This would make the scalars indeed point-like, a feature which is also observed in the unquenched case \cite{Maas:2013aia}. This is very different from gluons, which seem to have a dramatically large (imaginary) radius \cite{Maas:2011se,Cucchieri:2008qm,Maas:2007uv,Sternbeck:2017ntv,Huber:2018ned,Boucaud:2017obn,Binosi:2017rwj}, but very similar to ghosts \cite{Cucchieri:2006tf,Cucchieri:2008qm,Maas:2007uv,Cucchieri:2004sq,Sternbeck:2005re,Sternbeck:2012mf,Zafeiropoulos:2019flq,Maas:2011se,Huber:2018ned}.

Thus, the picture which emerges is that scalar particles are essentially point-like objects embedded into a background of extended gluons. Since their propagators show a mass scale, in contrast to the ghosts \cite{Maas:2011se}, they also are not mediating any long-range correlations. Hence, scalar matter behaves truly as essentially inert objects, with properties driven by the gauge dynamics.\\

\no{\bf Acknowledgments}\\

This work was supported by the DFG under grant numbers MA 3935/5-1, MA-3935/8-1 (Heisenberg program) and the FWF under grant number M1099-N16. Simulations were performed on the HPC clusters at the Universities of Jena and Graz. The author is grateful to the HPC teams for the very good performance of the clusters. The ROOT framework \cite{Brun:1997pa} has been used in this project.

\appendix

\section{Lattice setups}\label{a:ls}

The various lattice setups are listed in table \ref{tcgf}. The determination of the lattice spacings has been performed as in \cite{Maas:2014xma}.

\begin{longtable}{|c|c|c|c|c|c|c|c|}
\caption{\label{tcgf}Number and parameters of the configurations used, ordered by dimension, lattice spacing, and physical volume. In all cases $2(10N+100(d-1))$ thermalization sweeps and $2(N+10(d-1))$ decorrelation sweeps of mixed updates \cite{Cucchieri:2006tf} have been performed, and auto-correlation times of local observables have been monitored to be at or below one sweep. The number of configurations were selected such as to have a reasonable small statistical error for the renormalization constants determined in section \ref{s:ren}. The value $m_0$ denotes the value of the mass parameter in \prefr{covf}{cova} to yield a tree-level mass of 1 GeV. The other tree-level masses are obtained by multiplying or dividing this number by 10, or setting it to zero for tree-level mass zero.}\\
\hline
$d$	& $N$	& $\beta$	& $a$ [fm] & $a^{-1}$ [GeV]	& L [fm]	&  $m_0$	& config.	\endfirsthead
\hline
\multicolumn{8}{|l|}{Table \ref{tcgf} continued}\\
\hline
$d$	& $N$	& $\beta$	& $a$ [fm] & $a^{-1}$ [GeV]	& L [fm]	&  $m_0$	& config.	\endhead
\hline
\multicolumn{8}{|r|}{Continued on next page}\\
\hline\endfoot
\endlastfoot
\hline
2	& 92	& 6.23	& 0.228		& 0.863	& 21	& 1.159		& 2848	\cr
\hline
2	& 106	& 6.33	& 0.226		& 0.870	& 24	& 1.149		& 2752	\cr
\hline
2	& 80	& 6.40	& 0.225		& 0.875	& 18	& 1.143		& 1734	\cr
\hline
2	& 58	& 6.45	& 0.224		& 0.879	& 13	& 1.138		& 1734	\cr
\hline
2	& 18	& 6.55	& 0.222		& 0.886	& 4.0	& 1.129		& 1910	\cr
\hline
2	& 122	& 6.60	& 0.221		& 0.890	& 27	& 1.124		& 2158	\cr
\hline
2	& 34	& 6.64	& 0.221		& 0.893	& 7.5	& 1.120		& 1510	\cr
\hline
2	& 68	& 6.64	& 0.221		& 0.893	& 15	& 1.120		& 1746	\cr
\hline
2	& 10	& 6.68	& 0.220		& 0.895	& 2.2	& 1.117		& 1060	\cr
\hline
2	& 50	& 6.68	& 0.220		& 0.895	& 11	& 1.117		& 1774	\cr
\hline
2	& 26	& 6.72	& 0.219		& 0.898	& 5.7	& 1.113		& 1680	\cr
\hline
2	& 42	& 6.73	& 0.219		& 0.900	& 9.2	& 1.112		& 1880	\cr
\hline
2	& 106	& 8.13	& 0.198		& 0.994 & 21	& 1.006		& 2752	\cr
\hline
2	& 122	& 8.24	& 0.197		& 1.00	& 24	& 0.9990	& 2158	\cr
\hline
2	& 92	& 8.33	& 0.196		& 1.01	& 18	& 0.9933	& 1958	\cr
\hline
2	& 68	& 8.70	& 0.191		& 1.03	& 13	& 0.9708	& 1746	\cr
\hline
2	& 58	& 8.83	& 0.190		& 1.04	& 11	& 0.9632	& 1725	\cr
\hline
2	& 80	& 9.03	& 0.188		& 1.05	& 15	& 0.9519	& 2208	\cr
\hline
2	& 50	& 9.36	& 0.184		& 1.07	& 9.2	& 0.9341	& 1649	\cr
\hline
2	& 42	& 9.91	& 0.179		& 1.10	& 7.5	& 0.9066	& 1880	\cr
\hline
2	& 122	& 10.6	& 0.172		& 1.14	& 21	& 0.8752	& 2573	\cr
\hline
2	& 106	& 10.9	& 0.170		& 1.16	& 18	& 0.8625	& 2752	\cr
\hline
2	& 34	& 11.1	& 0.168		& 1.17	& 5.7	& 0.8543	& 1933	\cr
\hline
2	& 92	& 11.7	& 0.164		& 1.20	& 15	& 0.8312	& 2848	\cr
\hline
2	& 80	& 11.8	& 0.163		& 1.21	& 13	& 0.8275	& 1749	\cr
\hline
2	& 68	& 11.9	& 0.162		& 1.21	& 11	& 0.8239	& 1746	\cr
\hline
2	& 58	& 12.4	& 0.159		& 1.24	& 9.2	& 0.8065	& 1652	\cr
\hline
2	& 26	& 13.1	& 0.154		& 1.28	& 4.0	& 0.7838	& 1680	\cr
\hline
2	& 50	& 13.8	& 0.150		& 1.31	& 7.5	& 0.7629	& 1649	\cr
\hline
2	& 122	& 14.3	& 0.148		& 1.34	& 18	& 0.7490	& 2656	\cr
\hline
2	& 92	& 15.5	& 0.142		& 1.39	& 13	& 0.7185	& 2784	\cr
\hline
2	& 106	& 15.5	& 0.142		& 1.39	& 15	& 0.7185	& 3856	\cr
\hline
2	& 80	& 16.3	& 0.138		& 1.43	& 11	& 0.7001	& 1749	\cr
\hline
2	& 42	& 16.8	& 0.136		& 1.45	& 5.7	& 0.6893	& 1869	\cr
\hline
2	& 68	& 16.9	& 0.135		& 1.46	& 9.2	& 0.6872	& 1710	\cr
\hline
2	& 58	& 18.4	& 0.130		& 1.52	& 7.5	& 0.6578	& 1652	\cr
\hline
2	& 106	& 20.4	& 0.123		& 1.60	& 13	& 0.6239	& 2016	\cr
\hline
2	& 18	& 20.6	& 0.122		& 1.61	& 2.2	& 0.6208	& 1981	\cr
\hline
2	& 92	& 21.5	& 0.120		& 1.65	& 11	& 0.6074	& 2496	\cr
\hline
2	& 34	& 22.2	& 0.118		& 1.67	& 4.0	& 0.5974	& 1510	\cr
\hline
2	& 80	& 23.2	& 0.115		& 1.71	& 9.2	& 0.5841	& 1749	\cr
\hline
2	& 50	& 23.6	& 0.114		& 1.73	& 5.7	& 0.5791	& 1622	\cr
\hline
2	& 68	& 25.2	& 0.110		& 1.79	& 7.5	& 0.5600	& 1710	\cr
\hline
2	& 106	& 28.4	& 0.104		& 1.90	& 11	& 0.5269	& 3200	\cr
\hline
2	& 92	& 30.5	& 0.100		& 1.97	& 9.2	& 0.5082	& 2088	\cr
\hline
2	& 58	& 31.6	& 0.0983	& 2.00	& 5.7	& 0.4991	& 1650	\cr
\hline
2	& 42	& 33.6	& 0.0953	& 2.07	& 4.0	& 0.4838	& 1840	\cr
\hline
2	& 80	& 34.7	& 0.0938	& 2.10	& 7.5	& 0.4759	& 1749	\cr
\hline
2	& 106	& 40.4	& 0.0868	& 2.27	& 9.2	& 0.4406	& 2632	\cr
\hline
2	& 26	& 42.4	& 0.0847	& 2.33	& 2.2	& 0.4300	& 1680	\cr
\hline
2	& 68	& 43.2	& 0.0839	& 2.35	& 5.7	& 0.4260	& 1664	\cr
\hline
2	& 92	& 45.7	& 0.0816	& 2.42	& 7.5	& 0.4140	& 1924	\cr
\hline
2	& 50	& 47.4	& 0.0801	& 2.46	& 4.0	& 0.4064	& 1690	\cr
\hline
2	& 80	& 59.7	& 0.0713	& 2.76	& 5.7	& 0.3618	& 1800	\cr
\hline
2	& 106	& 60.5	& 0.0708	& 2.78	& 7.5	& 0.3593	& 2000	\cr
\hline
2	& 58	& 63.7	& 0.0690	& 2.86	& 4.0	& 0.3501	& 1566	\cr
\hline
2	& 34	& 72.3	& 0.0647	& 3.04	& 2.2	& 0.3285	& 1840	\cr
\hline
2	& 92	& 78.8	& 0.0620	& 3.18	& 5.7	& 0.3146	& 2304	\cr
\hline
2	& 122	& 80	& 0.03122	& 3.20	& 7.5	& 0.3122	& 2304  \cr
\hline
2	& 68	& 87.3	& 0.0589	& 3.35	& 4.0	& 0.2988	& 1736	\cr
\hline
2	& 106	& 104	& 0.0539	& 3.65	& 5.7	& 0.2736	& 1978	\cr
\hline
2	& 42	& 110	& 0.0524	& 3.76	& 2.2	& 0.2660	& 1568	\cr
\hline
2	& 80	& 120	& 0.0502	& 3.93	& 4.0	& 0.02546	& 1802	\cr
\hline
2	& 50	& 155	& 0.0441	& 4.47	& 2.2	& 0.2239	& 1525	\cr
\hline
2	& 92	& 159	& 0.0436	& 4.52	& 4.0	& 0.2211	& 1887	\cr
\hline
2	& 58	& 209	& 0.0380	& 5.19	& 2.2	& 0.1928	& 1652	\cr
\hline
2	& 106	& 211	& 0.0378	& 5.21	& 4.0	& 0.1919	& 2592	\cr
\hline
2	& 68	& 287	& 0.0324	& 6.08	& 2.2	& 0.1644	& 1710	\cr
\hline
2	& 80	& 398	& 0.0275	& 7.16	& 2.2	& 0.1396	& 1749	\cr
\hline
2	& 92	& 526	& 0.0239	& 8.24	& 2.2	& 0.1214	& 2784	\cr
\hline
\hline
3	& 48	& 3.35	& 0.230		& 0.858	& 11	& 1.166		& 1854	\cr
\hline
3	& 8	& 3.40	& 0.225		& 0.874	& 1.8	& 1.144		& 1500	\cr
\hline
3	& 54	& 3.43	& 0.223		& 0.884	& 12	& 1.131		& 3328	\cr
\hline
3	& 14	& 3.44	& 0.222		& 0.887	& 3.1	& 1.127		& 1800	\cr
\hline
3	& 20	& 3.46	& 0.220		& 0.894	& 4.4	& 1.119		& 1580	\cr
\hline
3	& 26	& 3.47	& 0.220		& 0.897	& 5.7	& 1.115		& 1420	\cr
\hline
3	& 36	& 3.47	& 0.220		& 0.897	& 7.9	& 1.115		& 1650	\cr
\hline
3	& 42	& 3.47	& 0.220		& 0.897	& 9.2	& 1.115		& 3296	\cr
\hline
3	& 32	& 3.48	& 0.219		& 0.900	& 7.0	& 1.111		& 1548	\cr
\hline
3	& 54	& 3.68	& 0.204		& 0.966	& 11	& 1.035		& 3328	\cr
\hline
3	& 36	& 3.82	& 0.195		& 1.01	& 7.0	& 0.9883	& 1650	\cr
\hline
3	& 48	& 3.86	& 0.192		& 1.03	& 9.2	& 0.9756	& 1854	\cr
\hline
3	& 42	& 3.92	& 0.189		& 1.04	& 7.9	& 0.9572	& 1725	\cr
\hline
3	& 32	& 4.10	& 0.178		& 1.10	& 5.7	& 0.9058	& 1458	\cr
\hline
3	& 54	& 4.25	& 0.171		& 1.15	& 9.2	& 0.8671	& 3328	\cr
\hline
3	& 26	& 4.28	& 0.169		& 1.16	& 4.4	& 0.8597	& 1420	\cr
\hline
3	& 42	& 4.33	& 0.167		& 1.18	& 7.0	& 0.8477	& 1725	\cr
\hline
3	& 48	& 4.38	& 0.165		& 1.20	& 7.9	& 0.8360	& 2976	\cr
\hline
3	& 36	& 4.52	& 0.159		& 1.24	& 5.7	& 0.8050	& 1650	\cr
\hline
3	& 20	& 4.60	& 0.155		& 1.27	& 3.1	& 0.7883	& 1580	\cr
\hline
3	& 54	& 4.83	& 0.147		& 1.34	& 7.9	& 0.7439	& 1768	\cr
\hline
3	& 48	& 4.84	& 0.146		& 1.35	& 7.0	& 0.7420	& 1800	\cr
\hline
3	& 32	& 5.09	& 0.138		& 1.43	& 4.4	& 0.6993	& 1522	\cr
\hline
3	& 42	& 5.15	& 0.136		& 1.45	& 5.7	& 0.6897	& 1725	\cr
\hline
3	& 60	& 5.29	& 0.132		& 1.50	& 7.9	& 0.6685	& 2496	\cr
\hline
3	& 54	& 5.36	& 0.130		& 1.52	& 7.0	& 0.6583	& 6512	\cr
\hline
3	& 14	& 5.39	& 0.129		& 1.53	& 1.8	& 0.6540	& 1800	\cr
\hline
3	& 36	& 5.64	& 0.122		& 1.61	& 4.4	& 0.6206	& 1650	\cr
\hline
3	& 66	& 5.74	& 0.120		& 1.64	& 7.9	& 0.6081	& 2178	\cr
\hline
3	& 26	& 5.76	& 0.119		& 1.65	& 3.1	& 0.6057	& 1420	\cr
\hline
3	& 48	& 5.78	& 0.119		& 1.66	& 5.7	& 0.6033	& 1725	\cr
\hline
3	& 54	& 6.41	& 0.106		& 1.87	& 5.7	& 0.5361	& 1976	\cr
\hline
3	& 42	& 6.45	& 0.105		& 1.88	& 4.4	& 0.5323	& 1725	\cr
\hline
3	& 32	& 6.91	& 0.0970	& 2.03	& 3.1	& 0.4925	& 1522	\cr
\hline
3	& 60	& 7.04	& 0.0950	& 2.07	& 5.7	& 0.4824	& 2106	\cr
\hline
3	& 48	& 7.27	& 0.0917	& 2.15	& 4.4	& 0.4653	& 3366	\cr
\hline
3	& 20	& 7.39	& 0.0900	& 2.19	& 1.8	& 0.4569	& 1580	\cr
\hline
3	& 66	& 7.67	& 0.0864	& 2.28	& 5.7	& 0.4384	& 2160	\cr
\hline
3	& 36	& 7.69	& 0.0861	& 2.29	& 3.1	& 0.4371	& 1650	\cr
\hline
3	& 54	& 8.08	& 0.0815	& 2.42	& 4.4	& 0.4139	& 1955	\cr
\hline
3	& 42	& 8.84	& 0.0739	& 2.67	& 3.1	& 0.3750	& 1725	\cr
\hline
3	& 26	& 9.38	& 0.0692	& 2.84	& 1.8	& 0.3515	& 1420	\cr
\hline
3	& 48	& 10.0	& 0.0646	& 3.05	& 3.1	& 0.3280	& 2046	\cr
\hline
3	& 54	& 11.1	& 0.0577	& 3.41	& 3.1	& 0.2931	& 1933	\cr
\hline
3	& 32	& 11.3	& 0.0566	& 3.48	& 1.8	& 0.2875	& 1704	\cr
\hline
3	& 36	& 12.7	& 0.0500	& 3.94	& 1.8	& 0.2539	& 1782	\cr
\hline
3	& 42	& 14.6	& 0.0432	& 4.57	& 1.8	& 0.2191	& 1701	\cr
\hline
3	& 48	& 16.6	& 0.0377	& 5.22	& 1.8	& 0.1914	& 1944	\cr
\hline
3	& 54	& 18.6	& 0.0335	& 5.88	& 1.8	& 0.1700	& 2624	\cr
\hline
3	& 60	& 20.6	& 0.0301	& 6.54	& 1.8	& 0.01529	& 1900	\cr
\hline
\hline
4	& 14	& 2.179	& 0.221		& 0.889	& 3.1	& 1.124		& 1420	\cr
\hline
4	& 10	& 2.181	& 0.220		& 0.894	& 2.2	& 1.119		& 1500	\cr
\hline
4	& 26	& 2.183	& 0.219		& 0.898	& 5.7	& 1.114		& 1505	\cr
\hline
4	& 22	& 2.185	& 0.218		& 0.902	& 4.8	& 1.109		& 1635	\cr
\hline
4	& 6	& 2.188	& 0.217		& 0.908	& 1.3	& 1.101		& 1110	\cr
\hline
4	& 18	& 2.188	& 0.217		& 0.908	& 3.9	& 1.101		& 1360	\cr
\hline
4	& 30	& 2.188	& 0.217		& 0.908	& 6.5	& 1.101		& 1933	\cr
\hline
4	& 30	& 2.241	& 0.190		& 1.03	& 5.7	& 0.9667	& 1910	\cr
\hline
4	& 26	& 2.252	& 0.185		& 1.06	& 4.8	& 0.9396	& 1650	\cr
\hline
4	& 32	& 2.266 & 0.178		& 1.10	& 5.7	& 0.9055	& 1733	\cr
\hline
4	& 22	& 2.268	& 0.177		& 1.11	& 3.9	& 0.9007	& 1512	\cr
\hline
4	& 18	& 2.279	& 0.172		& 1.14	& 3.1	& 0.8743	& 1355	\cr
\hline
4	& 30	& 2.305	& 0.160		& 1.23	& 4.8	& 0.8136	& 1702	\cr
\hline
4	& 14	& 2.311	& 0.158		& 1.25	& 2.2	& 0.7999	& 1420	\cr
\hline
4	& 26	& 2.328	& 0.150		& 1.31	& 3.9	& 0.7618	& 1628	\cr
\hline
4	& 22	& 2.349 & 0.141		& 1.40	& 3.1	& 0.7162	& 1484	\cr
\hline
4	& 10	& 2.376 & 0.130		& 1.52	& 1.3	& 0.6600	& 1500	\cr
\hline
4	& 30	& 2.376 & 0.130		& 1.52	& 3.9	& 0.6600	& 1760	\cr
\hline
4	& 18	& 2.395	& 0.123		& 1.61	& 2.2	& 0.6222	& 1792	\cr
\hline
4	& 26	& 2.403	& 0.120		& 1.65	& 3.1	& 0.6067	& 1625	\cr
\hline
4	& 30	& 2.448 & 0.103		& 1.91	& 3.1	& 0.5246	& 1633	\cr
\hline
4	& 22	& 2.457	& 0.100		& 1.96	& 2.2	& 0.5092	& 1749	\cr
\hline
4	& 14	& 2.480	& 0.0929	& 2.12	& 1.3	& 0.4714	& 1420	\cr
\hline
4	& 26	& 2.507	& 0.0847	& 2.33	& 2.2	& 0.4299	& 1611	\cr
\hline
4	& 30	& 2.548	& 0.0734	& 2.68	& 2.2	& 0.3726	& 1750	\cr
\hline
4	& 18	& 2.552	& 0.0724	& 2.72	& 1.3	& 0.3674	& 1360	\cr
\hline
4	& 32	& 2.566	& 0.0689	& 2.86	& 2.2	& 0.03496	& 1950  \cr
\hline
4	& 22	& 2.609	& 0.0591	& 3.33	& 1.3	& 0.3001	& 1732	\cr
\hline
4	& 26	& 2.656	& 0.0501	& 3.93	& 1.3	& 0.2543	& 2049	\cr
\hline
4	& 30	& 2.698	& 0.0434	& 4.54	& 1.3	& 0.2204	& 1838	\cr
\hline
\end{longtable}

\bibliographystyle{bibstyle}
\bibliography{bib}


\end{document}